\documentclass[12pt]{article}
\pdfoutput=1
\usepackage{bbm}
\usepackage{color}
\usepackage{authblk}
\usepackage{latexsym}
\usepackage{epsfig,amssymb,euscript, mathrsfs,cite}
\usepackage{amsmath}
\usepackage{mathrsfs} 
\usepackage{enumerate}
\definecolor{MyDarkBlue}{rgb}{0.15,0.15,0.45}
\usepackage[linktocpage=true]{hyperref}
\hypersetup{
colorlinks=true,
citecolor=MyDarkBlue,
linkcolor=MyDarkBlue,
urlcolor=MyDarkBlue,
pdfauthor={},
pdftitle={},
pdfsubject={hep-th}
}

\newsavebox{\ns}
\newsavebox{\dbrane}
\newsavebox{\dbshort}

\def\be{\begin{equation}}
\def\ee{\end{equation}}
\def\bea{\begin{eqnarray}}
\def\eea{\end{eqnarray}}

\newcommand{\nn}{\notag \\}

\def\cO{{\mathcal O}}
\def\eq#1 { \begin{equation} #1 \end{equation} }

\newcommand{\bb}{\bar{\beta}}

\newcommand{\cn}{h_0}

\newcommand{\dd}{\mathrm{d}}

\newlength{\sswidth}

\newcommand\s{\mathcal{\sigma}}
\newcommand\n{\mathcal{\beta}}

\numberwithin{equation}{section}       

\begin{document}

\begin{titlepage}

\vfill

\begin{flushright}
Imperial/TP/2018/JG/05\\
\end{flushright}

\vfill

\begin{center}
   \baselineskip=16pt
   {\Large\bf Spatially modulated and supersymmetric\\ deformations of ABJM theory}
  \vskip 1.5cm
Igal Arav, Jerome P. Gauntlett\\
Matthew M. Roberts and Christopher Rosen\\
     \vskip .6cm
             \begin{small}\vskip .6cm
      \textit{Blackett Laboratory, 
  Imperial College\\ London, SW7 2AZ, U.K.}
        \end{small}\\
                       \end{center}
\vfill

\begin{center}
\textbf{Abstract}
\end{center}
\begin{quote}
 We construct supersymmetric solutions of $D=11$ supergravity, preserving 1/4 of the supersymmetry, that are holographically
dual to ABJM theory which has been deformed by spatially varying mass terms depending
on one of the two spatial directions. 
We show that the BPS equations reduce to 
the Helmholtz equation on the complex plane leading to rich classes of new solutions.
In particular, the construction gives rise to infinite classes of supersymmetric boomerang RG flows, as well as generalising a known 
Janus solution. 
\end{quote}

\vfill

\end{titlepage}

\newpage

\section{Introduction}\label{sec:intro}

One of the canonical examples of the AdS/CFT correspondence is provided by ABJM theory \cite{Aharony:2008ug}. This is a three-dimensional Chern-Simons theory coupled to matter
with gauge group $U(N)_q \times U(N)_{-q}$, where $q$ labels the Chern-Simons level on each factor. It has manifest $\mathcal{N}=6$ supersymmetry, with an  
enhancement to $\mathcal{N}=8$ when $q=1,2$. The theory describes the low energy dynamics of $N$ M2 branes on $\mathbb{C}^4/\mathbb{Z}_q$. In the large $N$ and strongly coupled limit, the physics of the ABJM theory is captured holographically by $D=11$ supergravity on $AdS_4\times S^7/\mathbb{Z}_q$.

There is an interesting mass deformation of ABJM theory which breaks the conformal symmetry but
continues to preserve 
$\mathcal{N}=6$ supersymmetry \cite{Gomis:2008vc,Hosomichi:2008jb}. Furthermore, the
gravity dual was identified in \cite{Lin:2004nb}, extending \cite{Pope:2003jp,Bena:2004jw}.
In an interesting recent development, it was shown that one can deform ABJM theory
with mass terms that are spatially varying in one of the two spatial directions of the boundary theory while
preserving two-dimensional Poincar\'e invariance and $\mathcal{N}=3$ supersymmetry \cite{Kim:2018qle}.

The field content of ABJM  includes bosons, $Y^A$, and fermions, $\psi_A$ transforming under 
the $SU(4)\times U(1)_b$ global symmetry in the $\bf{4}_0$ and $\bar{\bf{4}}_0$, respectively. 
The mass deformation of the ABJM theory of \cite{Kim:2018qle} involves a correlated combination of mass terms for both fermions and bosons.
It includes the terms\footnote{There is also a term $m^2(\s)\mathrm{Tr} \left(Y^A Y^\dagger_A\right)$,
which is an unprotected operator and hence is not visible from the supergravity point of view.}
\begin{align}\label{kimetal}
\Delta\mathcal{L}=m'(\s)\mathcal{O}_\mathcal{X}^{\Delta=1} +m(\sigma)\mathcal{O}_\mathcal{Y}^{\Delta=2}+\dots\,,
\end{align}
where $m(\sigma)$ is an arbitrary function depending on one of the two spatial directions, $\s$, and the two operators are given by
\begin{equation}\label{eq:150}
\mathcal{O}_\mathcal{X}^{\Delta = 1} \sim M_{A}{}^{B}\mathrm{Tr}\big(Y^A Y^\dagger_B \big), \qquad 
\mathcal{O}_\mathcal{Y}^{\Delta = 2} \sim M_{A}{}^{B}\mathrm{Tr}\big(\psi^\dagger\,^A\psi_B +\frac{8\pi}{q}Y^CY^\dagger_{[C}Y^A Y^\dagger_{B]} \big)\,,
\end{equation}
with $M_A\,^B = \mathrm{diag}(1,1,-1,-1)$.

In this paper we will investigate the holographic duals of these deformations by constructing new solutions
of $D=11$ supergravity which generically preserve 1/4 of the supersymmetry. We will use the fact that
$\mathcal{N}=8$ $SO(8)$ gauged supergravity in $D=4$ arises as a consistent KK truncation of $D=11$ supergravity on the seven-sphere. Thus, any
solution of this $D=4$ theory can be uplifted on the seven sphere to obtain an exact solution of $D=11$ supergravity.
The two operators appearing in \eqref{eq:150}
are dual to fields $\mathcal{X}$, $\mathcal{Y}$ that lie within the $SO(4)\times SO(4)\subset SO(8)$ invariant sector of the $\mathcal{N}=8$ theory. As a consequence,
we can construct new solutions using a simple $D=4$ theory of gravity coupled to a single complex scalar field $z=\mathcal{X}+i\mathcal{Y}$.

A special example of such solutions was recently presented in \cite{Gauntlett:2018vhk}. Indeed 
a ``Susy Q solution" was constructed that is dual to a spatially periodic mass deformation with 
$m(\sigma)$ having a single Fourier mode e.g. $m(\sigma)\sim \sin k\sigma$. 
This construction arose out of a broader programme of investigating CFTs which 
have been deformed by spatially periodic sources.
The dual gravitational solutions,
known as holographic lattices \cite{Horowitz:2012ky}, provide a natural holographic framework for
studying transport of heat and charge, since the spatially varying sources provide a
mechanism for momentum to dissipate. Holographic lattices are also a powerful framework for
discovering novel holographic ground states with potential applications to condensed matter systems
\cite{Donos:2012js,Donos:2014uba,Gouteraux:2014hca}. The Q lattice construction of
\cite{Donos:2013eha} exploits a global symmetry in the bulk in order to construct holographic lattices by solving a system of ODEs rather than
PDEs. Within the context of the $SO(4)\times SO(4)$ invariant sector of $\mathcal{N}=8$ $D=4$ gauged supergravity,
Q lattice solutions that preserve supersymmetry, hence the name Susy Q, were constructed in \cite{Gauntlett:2018vhk}. 
Associated with the preservation of supersymmetry, it was shown that in the dual deformed ABJM theory the spatially averaged energy 
density is exactly zero.
Another interesting feature of the Susy Q solutions is that they describe boomerang RG flows \cite{Donos:2017ljs}, flowing from 
the $AdS_4$ vacuum in the UV to exactly the
same $AdS_4$ vacuum in the IR \cite{Chesler:2013qla}.

Another special class of known solutions is the $D=11$ Janus-type solutions of \cite{DHoker:2009lky,Bobev:2013yra}, which
includes the special case with $m(\sigma)\sim \delta(\sigma)$. While for generic $m(\sigma)$ a two-dimensional Poincar\'e invariance
is preserved, the Janus solutions are distinguished in preserving two-dimensional conformal invariance. The dual gravitational
solutions can therefore be constructed by foliating the spacetime with $AdS_3$ factors. The solutions
in \cite{DHoker:2009lky} were constructed directly in $D=11$ supergravity, while the solutions in \cite{Bobev:2013yra} of relevance here 
were constructed using the $SO(4)\times SO(4)$ invariant sector of $\mathcal{N}=8$ $D=4$ gauged supergravity.

In this paper we will generalise these solutions by
constructing infinite classes of new supersymmetric solutions of $D=11$ supergravity
that are associated with more general spatial deformations of ABJM theory. 
Remarkably, 
the entire problem reduces to solving
the Helmholtz equation\footnote{More precisely, the Helmholtz equation with imaginary value for the wave-number, also known as the ``homogeneous screened Poisson equation"; see\eqref{helm}} on the complex plane.  For example, the Susy Q solution 
is associated with a solution to the Helmholtz equation on a circular disc that vanishes
on the edge of the disc and with a delta function source at the centre of the disc. This construction
can simply be extended to construct infinite classes of solutions with periodic spatial deformations
that are also boomerang RG flows.

The plan of the paper is as follows. In section \ref{sec2} we present the BPS equations and show that they reduce to
solving the Helmholtz equation on the complex plane. In order to get a quick feel for the BPS equations
we show how to recover the dielectric RG flow solution, the Janus solution and the Susy Q solution in section \ref{sec3}.
Section \ref{gencoms} contains some general analysis of the BPS equations and section \ref{sec5} concentrates on solutions
associated with boomerang RG flows. 
Section \ref{sec6} concludes with some discussion. 
Appendix \ref{appa} contains some details on the derivation of the BPS equations, in a more general setting,
and appendix \ref{sec:HolographicOnePointFunctions} has some details on the holographic calculation of the sources and VEVs in the dual ABJM theory. Appendix \ref{sec:dist} contains some details of how to deal with solutions with distributional sources as in, for example, the Janus solution.
Finally, appendix \ref{simpsol} discusses some additional simple analytic solutions to the BPS equations, including a discussion
of the sources and VEVs for the Janus solution.

\section{BPS equations}\label{sec2}
Our starting point is the $SO(4)\times SO(4)$ invariant sector of $\mathcal{N}=8$ gauged supergravity in $D=4$. 
The bosonic content 
consists of a metric coupled to a complex scalar field, $z$, which parametrises the Poincar\'e disc with $|z|<1$ and
action given by
\begin{equation}\label{eq:Sg2bulk}
S= \int\dd^4 x \sqrt{-g}\Big(R - \frac{2}{(1-|z|^2)^2}\partial_\mu z \partial^\mu \bar{z}
+\frac{2(3-|z|^2)}{1-|z|^2}\Big)\,.
\end{equation}
Any solution to the equations of motion gives rise to an exact solution of $D=11$ supergravity
after uplifting on the seven sphere using the formulae in \cite{Cvetic:1999au}. By taking a $\mathbb{Z}_q$ quotient
of the seven-sphere one obtains solutions relevant to ABJM theory.

We are interested in spatial deformations of ABJM theory that depend on one of the two spatial coordinates and
maintains Poincar\'e invariance in the remaining two directions, which we label with coordinates $(t,y)$. 
The ansatz we shall consider is therefore given by
\begin{align}\label{Susyqansatz}
ds^2&=e^{2A}(-dt^2+dy^2)+e^{2V} dx^2 +N^2dr^2\,,
\end{align}
with $A$, $V$, $N$, and the scalar field $z$ all functions of $(x,r)$.
There is some residual coordinate freedom, involving 
$(x,r)$ and maintaining the form of the ansatz \eqref{Susyqansatz}, that we return to below.
The $AdS_4$ vacuum solution
has $e^A=e^V=r$, $N=1/r$ and $z=0$ and uplifts to the maximally supersymmetric $AdS_4\times S^7$ solution. 
We also note that the Susy Q solutions constructed in \cite{Gauntlett:2018vhk} have $z=f(r)e^{ikx}$ with $A$, $V$ and
$N$ all independent of $x$. If, in addition, we set $k=0$ and $A=V$ then we recover the ansatz for the
1/2 supersymmetric and Poincar\'e invariant dielectric RG flows of \cite{Pope:2003jp}.

We demand that the solutions preserve (at least) 1/4 of the supersymmetry. Some details
of this analysis is presented in appendix \ref{appa}, and here we just summarise the main results.
In the $SO(4)\times SO(4)\subset SO(8)$ sector we can decompose the Majorana supersymmetry parameters
into $\epsilon^a_{(M)}$ and $\epsilon^s_{(M)}$, transforming in the ${\bf (4,1)}$ and ${\bf (1,4)}$, respectively.
Using the orthonormal frame $(e^{\hat{t}},e^{\hat{x}},e^{\hat{y}},e^{\hat{r}})= 
(e^A dt, e^V dx, e^A dy, Ndr)$, the Killing spinors 
satisfy the following projections
\begin{align}
\Gamma^{\hat{t}\hat{y}}\epsilon^a_{(M)}&= +\epsilon^a_{(M)}\,,\qquad\left[-\sin\xi\,\Gamma^{\hat{x}}+\cos \xi\,\Gamma^{\hat{r}}\right]\epsilon^a_{(M)} = -\epsilon^a_{(M)} \,,\nn
\Gamma^{\hat{t}\hat{y}}\epsilon^s_{(M)}&= -\epsilon^s_{(M)}\,,\qquad 
\left[-\sin\xi\,\Gamma^{\hat{x}}+\cos \xi\,\Gamma^{\hat{r}}\right]\epsilon^s_{(M)} = -\epsilon^s_{(M)}\,,
\end{align}
with the functional dependence given in \eqref{fundepks}.
The associated BPS equations are given by
\begin{align}\label{bpsderxi1}
(1-|z|^2)^{1/2}\left(N^{-1}\partial_r A-ie^{-V}\partial_x A\right)&=e^{i\xi}\,,\nn
\left(ie^{-V}\partial_x z+N^{-1}\partial_r z\right)e^{i\xi}+(1-|z|^2)^{1/2}z&=0\,,\nn
e^{-V}\partial_x\xi-N^{-1}\partial_r V-\frac{N^{-1}\partial_r|z|^2}{2(1-|z|^2)}+(1-|z|^2)^{1/2}\cos\xi&=0\,,\nn
N^{-1}\partial_r\xi+e^{-V}N^{-1}\partial_x N+\frac{e^{-V}\partial_x|z|^2}{2(1-|z|^2)}+(1-|z|^2)^{1/2}\sin\xi&=0\,.
\end{align}
After considering the integrability condition for the first equation we can deduce that one of the last two equations
is not independent from the rest. We thus have five real equations for six real variables, 
$A,V,N,\xi$ and $z$; this redundancy is connected with residual coordinate freedom 
for the ansatz \eqref{Susyqansatz} that we mentioned above.
We also note that the BPS equations  for the Susy Q solutions given in \cite{Gauntlett:2018vhk} can be recovered by setting $\xi=0$, $\partial_x\epsilon=\partial_x A=\partial_x V=\partial_x N=0$ and $z=f(r)e^{ikx}$. 

At this stage we note that three choices of gauge suggest themselves, and 
in fact we will utilise all three of them. The first is the Fefferman-Graham gauge with
$N=1/r$, which was used in the numerical construction of the Susy Q solutions in \cite{Gauntlett:2018vhk}; we use it here
in appendix \ref{sec:HolographicOnePointFunctions} to carry out the holographic calculation of one point functions. The second is when 
the metric transverse to the $t,y$ directions is conformally flat, which is achieved when $N=e^V$ and we will use this shortly. 
The third gauge has $e^{A}=1/r$ which, from \eqref{bpsderxi1}, 
has the feature that $\xi=0$.

We continue now in the gauge
$N=e^V$ so that
\begin{align}\label{Susyqansatz2}
ds^2&=e^{2A}(-dt^2+dy^2)+e^{2V}dwd\bar w\,,
\end{align}
where we also introduced the complex coordinate $w\equiv r+i x$.
After defining the $(1,0)$ form 
\begin{align}\label{Bdef}
B\equiv \frac{e^{V+i\xi}}{2(1-|z|^2)^{1/2}}dw\,,
\end{align}
we can recast the BPS equations
\eqref{bpsderxi1} in the following remarkably compact form
\begin{align}\label{susygenint}
\partial A&=B\,,\nn
\bar \partial z&=-z(1-|z|^2)\bar B\,,\nn
\bar \partial B&=-(1-|z|^2)B\wedge \bar B\,,
\end{align}
where we have used the holomorphic and anti-holomorphic 
exterior derivatives $\partial$ and $\bar\partial$, respectively.
These equations can be further simplified as follows. The second and third
equations in \eqref{susygenint} imply that $\bar \partial(zB)=0$ and hence locally we have
\begin{align}\label{beecee}
zB=C(w)dw\,,
\end{align}
where the component $C(w)$ is an {\it arbitrary} holomorphic local function of $w$. 
We next define
\begin{align}\label{aichdef}
H=e^{-A}\,,
\end{align}
with $H=H(w,\bar w)$. If we now substitute \eqref{beecee} into the first equation
in \eqref{susygenint} and then take a derivative with respect to $\bar w$,
we deduce that
\begin{align}\label{helm1}
\partial_{w\bar w} H -|C|^2H=0\,.
\end{align}
Note that $\partial_{w\bar w}=1/4\nabla^2$ where $\nabla^2$ is the Laplacian for the flat transverse space with metric $dw d\bar w=dx^2+dr^2$. In fact, remarkably, 
for a given function $C$ and solution $H$
to \eqref{helm1} we can recover the full supersymmetric solution via
\begin{align}\label{Susyqansatz3}
ds^2&=H^{-2}\left[-dt^2+dy^2+((\nabla H)^2-4H^2|C|^2)dw d\bar w\right]\,,\nn
z&=-\frac{C}{\partial_w \ln H},
\end{align}
and $(\nabla H)^2=4\left|\partial_w H\right|^2$.
The holomorphic function $C(w)$ is associated with a
remaining gauge freedom corresponding to the holomorphic change of coordinates $w=w(\tilde w)$.

At this point we pause to observe that the $AdS_4$ vacuum solution with $z=0$ is obtained
by taking $C=0$ and solving \eqref{helm1} with $H=r$. However, since we will be interested in constructing
solutions with $z\ne 0$, we find it convenient to not choose $C=0$, but instead work in the gauge with
\begin{align}\label{chalf}
C(w)=\frac{1}{2}\,.
\end{align}
Indeed in this gauge the whole problem boils down to solving the Helmholtz equation for 
$H=H(w,\bar w)$ on the complex plane with flat metric $dw d\bar w$:
\begin{align}\label{helm}
\nabla^2 H -H=0\,.
\end{align}
To summarise, for any solution of the Helmholtz equation \eqref{helm} on the complex plane, the following solution preserves 1/4 of the supersymmetry
\begin{align}\label{Susyqansatz4}
ds^2&=H^{-2}\left[-dt^2+dy^2+((\nabla H)^2-H^2)dwd\bar w\right]\,,\nn
z&
=-\frac{1}{2\partial_w \ln H}\,.
\end{align}
The function $\xi$ appearing in Killing spinors can be obtained from \eqref{Bdef} 
using $B=1/(2z)dw$.

Since $H$ satisfies a linear equation it is now possible to construct rich classes of new solutions some of which will be explored in
the sequel. 
From \eqref{aichdef} we demand $H>0$ and we note that $H=0$ is associated with the holographic boundary.
We also recall from
\eqref{eq:Sg2bulk} that the boundary of the scalar field space is when $|z|=1$ and hence for a physical solution we require that
\begin{align}
|z|<1\qquad\Leftrightarrow \qquad H^2<(\nabla H)^2\,,
\end{align}
a condition that is also clearly required for the metric in \eqref{Susyqansatz4} to have the correct signature.
In particular, $H$ cannot have a critical point with $\nabla H=0$. 
We also notice that if we scale any solution $H$ by a constant then we obtain the same solution after scaling the $(t,y)$ coordinates.

For any given solution to the BPS equations we can obtain a new solution by simply multiplying $z$ by a constant phase $e^{i\zeta}$.
While this is somewhat trivial from the $D=4$ point of view, it has non-trivial consequences for the interpretation of the $D=11$ solutions in terms of membranes and fivebranes as first emphasised in \cite{Pope:2003jp}.

By solving the Helmholtz equation to construct BPS solutions, one often encounters singularities. In the context of holography
a criterion for determining whether or not a singularity in the IR region of a stationary geometry has a physical interpretation was presented in \cite{Maldacena:2000mw}. In the strongest form one demands that the $g_{tt}$ component of the metric should not increase as one
approaches the singularity, where $\partial_t$ is the time-like Killing vector. Importantly, in the present setting $g_{tt}$ refers to the component of the $D=11$ metric. After uplifting our solutions on $S^7$ using \cite{Cvetic:1999au} 
we find
\begin{equation}
g_{tt} = \frac{1}{H^2}{(Z\tilde{Z} )^{1/3}}=\frac{(Z\tilde{Z} )^{1/3}}{4|z|^2 |\partial_w H|^2}\,,
\end{equation}
where
\begin{equation}
Z = \frac{|1+z|^2}{1-|z|^2}\cos^2\nu +\sin^2\nu\,, \qquad \tilde{Z} = \frac{|1-z|^2}{1-|z|^2}\sin^2\nu+\cos^2\nu\,,
\end{equation}
and the coordinate $\nu$ with $0\le\nu\le\pi/2$ parametrises the interval in the foliation of $S^7$ by the two $S^3$'s in the uplift.
We see, for example, that an IR singularity with $|z|\to 1$ will be an unphysical singularity unless $|\partial_w H|^2$ 
diverges more rapidly than  $(Z\tilde{Z})^{1/3}$ as the singularity is approached for any value of $\nu$. 
In particular, if $|z|\to 1$ and $H$ remains finite then we have a bad singularity.

\section{Some known solutions}\label{sec3}
Before discussing new solutions we first recover some known solutions.

\subsection{Dielectric RG flow solution: half plane}
Recall that this solution is associated with a mass deformation of ABJM theory that preserves three-dimensional Poincar\'e invariance. To recover this solution, first given in \cite{Pope:2003jp}, 
we solve the Helmholtz equation on the upper half plane, parametrised with coordinates $(x,r)$,
with $H=H(r)$ and impose that $H=0$ on the $x$-axis.
Specifically we solve the Helmholtz equation using
\begin{align}
H= \sinh r\,,
\end{align}
to find that the full supersymmetric solution \eqref{Susyqansatz4} is then given by
\begin{align}\label{dielectric}
{}ds^2&=\frac{1}{\sinh^2 r}(-dt^2+dy^2+dx^2+dr^2)\,,\nn
z&=-\tanh r\,.
\end{align}
The $AdS_4$ boundary is located at $r\to 0$, where $H=0$. As $r\to\infty$
the solution becomes singular and it is a good singularity by the criteria discussed at the end of
the last subsection.

If we do the coordinate change $\sinh r=(2\mu)\tilde r/(\tilde r^2-\mu^2)$, where $\mu$ is constant, then we precisely obtain the
solution in the form given in eq. (3.11) of \cite{Gauntlett:2018vhk} after rescaling the $t,x,y$ coordinates by $2\mu$
and noting that in any supersymmetric solution we can multiply $z$ by a constant phase. 
All values of $\mu$ are physically equivalent.

\subsection{Janus solution: infinite strip}\label{janusfirst}
We next consider the supersymmetric Janus solution constructed in \cite{DHoker:2009lky,Bobev:2013yra}.
We now solve the Helmholtz equation on a strip in the $r, x$ plane with $H$ vanishing at $r=0$ and at $r=\pi/c$, for some constant $c>0$. We also demand that $H$ vanishes at $x\to-\infty$. We therefore take
\begin{align}
H=\sin(cr) e^{ \sqrt{1+c^2}x}\,,
\end{align}
leading to the solution 
\begin{align}\label{janus}
{}ds^2&=\frac{1}{\sin^2(c r) e^{2\sqrt{1+c^2} x}}(-dt^2+dy^2)+\frac{c^2}{\sin^2(cr)}(dx^2+dr^2)\,,\nn
z&=-\frac{c\cot(cr)+i\sqrt{1+c^2}}{c^2\cot^2(c r)+1+c^2}\,.
\end{align}
The $AdS_4$ boundary, where $H=0$, has three connected components located at $r=0$, $r=\pi/c$ and $x\to-\infty$.

If we now do the coordinate change $e^{\mu}=\tan(cr/2)$ then the solution can be written in a form
in which there is a manifest $AdS_3$ factor, parametrised by $(t,y,x)$:
\begin{align}\label{Susyqansatz5}
{}ds^2&=\cosh^2\mu[e^{-2(1+c^2)^{1/2} x}(-dt^2+dy^2)+c ^2dx^2]+d\mu^2\,,\nn
z&=\frac{c \sinh \mu-i\sqrt{1+c^2}}{c^2\cosh^2 \mu+1}\,,
\end{align}
This is in agreement with the solution in section 5.2 of \cite{Bobev:2013yra}.

In appendix \ref{simpsol} we extend the discussion of \cite{DHoker:2009lky,Bobev:2013yra} to 
extract out the one point functions for the stress tensor and scalar operators, using the results presented in section \ref{gencoms}.

\subsection{Susy Q: circular disc}
We next recover the Susy Q solution that was constructed numerically in \cite{Gauntlett:2018vhk}.
We now solve the Helmholtz equation inside a disc, with circular symmetry and 
which vanishes on the boundary of the disc. Thus, we introduce polar coordinates 
$w=\rho e^{i\theta}$, take $H=H(\rho)$ and demand that $H(c)=0$ for some $c>0$.
The solution to the Helmholtz equation can then be we written in terms of modified Bessel
functions as
\begin{align}\label{susyq1}
H=I_0(c) K_0\left(\rho\right) - K_0(c) I_0\left(\rho\right)\,,
\end{align}
with
\begin{align}\label{susyq2}
ds^2&=H^{-2}\left[-dt^2+dy^2+((\partial_\rho H)^2-H^2)(d\rho^2+\rho^2d\theta^2)\right]\,,\nn
z&=-\frac{1}{\partial_\rho \ln H}e^{i\theta}\,.
\end{align}
To recover the solution as in \cite{Gauntlett:2018vhk} one should set
$\theta =kx$ and $\rho= c e^{k r}$.

The UV $AdS_4$ boundary is located along the boundary of the disc, $\rho=c$, where
$H=0$. As explained in \cite{Gauntlett:2018vhk}, these solutions are dual to boomerang RG flows which approach the same $AdS_4$ vacuum in the IR. In the parametrisation here, the IR $AdS_4$
region is located at $\rho=0$ where $H$ develops a logarithmic singularity, 
$H\sim -\cn\log \rho$, with $\cn=I_0(c)$,
associated with a delta function appearing in the right hand side of the Helmholtz equation
$\nabla^2H-H=-2\pi \cn\delta(w,\bar w)$. 
We will discuss these solutions further and generalise them in section \ref{sec5}.

\section{General comments on the new solutions}\label{gencoms}

\subsection{Identifying the holographic boundary}\label{identbdy}
As in the specific examples just considered, the UV $AdS_4$ boundary sits on a curve in 
the complex plane defined by $H(w,\bar w)=0$. It is helpful
to consider a coordinate system, $\s,\n$ defined in some neighbourhood of the $H=0$ boundary curve 
with $\s$ parametrising arc length in the holographic boundary metric (which we note is not equal to the coordinate $x$, in general)
and $\n$ parametrising curves of constant $H$ i.e. $\n=H$. To do this
we first introduce $\s$ via
\begin{equation}\label{sigdef}
2 |\partial_w H| \left| \frac{dw}{d\s} \right| = 1\,.
\end{equation} 
On the $H=0$ curve we have $dH/d\s=0$ . Thus $(\partial H/\partial w)(dw/d\s)$ is imaginary on this curve
and so we can write the tangent vector as
\begin{equation}\label{vec1}
\frac{dw}{d\s} = - \frac{i}{2} \frac{1}{\partial_w H} =i\frac{z}{H}\,.
\end{equation}
We can now introduce another coordinate $\n$, with $\n=0$ on the $H=0$ curve, and determined away from this curve by 
demanding that
\begin{equation}
\frac{dw}{d\n} = \frac{1}{2 \partial_w H}\,.
\end{equation}
Notice, in particular, that this vector is orthogonal to the tangent vector \eqref{vec1} on the $H=0$ curve. 
Starting at $H=0$, we then define $\s$ away from the boundary to be constant on each of these integral curves.
We thus have constructed a coordinate system which is well-defined in some neighbourhood of the $H=\n=0$ curve, 
satisfying
\begin{equation}
\frac{\partial w}{\partial \s} = - \frac{i}{2} \frac{1}{\partial_w H}, \qquad \text{for $\n=0$}\,,
\end{equation}
on the boundary and
\begin{equation}
\frac{\partial w}{\partial \n} = \frac{1}{2 \partial_w H}\,,
\end{equation}
throughout the neighbourhood. From the above conditions we have
\begin{equation}
\frac{\partial H}{\partial \n} = \partial_w H \frac{\partial w}{\partial \n} + \partial_{\bar{w}} H \frac{\partial \bar{w}}{\partial \n} = 1\,,
\end{equation}
and since $H(\s,\n=0) =0$, it follows that $H=\n$ as claimed above.

Proceeding with the coordinates $(\s,\n)$, we now define $ I(\s,\n) \equiv 2i \partial_w H \frac{\partial w}{\partial \s} $, with $I(\s,\n=0) = 1$. 
Consider the one-form
\begin{equation}
J \equiv \partial_w H dw  = - \frac{H}{2z} dw\,,
\end{equation}
which we can also write as
\begin{equation}
\label{eq:JOneFormInSNCoordinates}
J = - \frac{i}{2} I d\s + \frac{1}{2} d\n\,.
\end{equation}
Since $ dH = d\n = J + \bar{J} $ it immediately follows that $I$ must be real.
Taking the exterior derivative of $J$ and using the Helmholtz equation we have:
\begin{equation}
dJ = \partial_{\bar w} \partial_w H d\bar{w} \wedge dw = \frac{1}{4} H d\bar{w} \wedge dw = \frac{|z|^2}{H} \bar{J}\wedge J\,.
\end{equation}
Substituting \eqref{eq:JOneFormInSNCoordinates} into the left hand side of this equality we 
obtain the following equation for the function $I$
\begin{equation}
\label{eq:IEqInSNCoordiantes}
\frac{\partial I}{\partial \n} = \frac{|z|^2}{H} I\,.
\end{equation}
The metric \eqref{Susyqansatz4} can now be written as follows:
\begin{equation}
\begin{split}
ds^2 
&= \n^{-2} \left[ -dt^2 + dy^2 + (1 - |z|^2) (I^2 d\s^2 + d\n^2) \right]\,,
\end{split}
\end{equation}
and we notice that this is now in the third gauge mentioned just below equation \eqref{bpsderxi1}.

If we expand around $\n=0$ we can write  $z=z_1 \n + z_2 \n^2 + O(\n^3) $ and from \eqref{eq:IEqInSNCoordiantes} we get $I = 1 + \frac{1}{2} |z_1|^2\n^2 + \frac{2}{3} \operatorname{Re}(z_1 \bar{z_2}) \n^3 + O(\n^4) $. Therefore, the metric can be written
\begin{equation}\label{expnigbet}
\begin{split}
ds^2 &=\beta^{-2} \left[ -dt^2 + dy^2 
+ (1 - \tfrac{2}{3} \operatorname{Re}(z_1 \bar{z_2}) \n^3 + O(\n^4))d\s^2 \right. \\
&\qquad\qquad \left. + \left(1-  |z_1|^2\n^2 - 2 \operatorname{Re}(z_1 \bar{z_2}) \n^3 + O(\n^4)\right) d\n^2 \right] .
\end{split}
\end{equation} 
We also observe, from the relation $ dw = -\frac{2z}{H} J$, that
\begin{equation}
d \left( \frac{z}{H} J \right) = 0 \quad \Rightarrow \quad
d\left(\frac{z}{H}\right)\wedge J + z \frac{|z|^2}{H^2} \bar{J}\wedge J =0\,.
\end{equation} 
Evaluating at $\n=0$ the second term in the last expression vanishes and we obtain the following relation between $z_1$ and $z_2$
\begin{equation}
z_2 = i \frac{dz_1}{d\s}.
\end{equation}
Thus, the expansion of the full solution is specified by $z_1$. Finally, we note that
from \eqref{vec1} we have \
\begin{align}\label{dwsz1}
\frac{dw}{d\s} = i z_1\,.
\end{align}
Equation \eqref{dwsz1} may be generalised to cases in which $z_1(\sigma)$ is a distribution, as 
discussed in appendix \ref{sec:dist}.

\subsection{Holographic one point functions}\label{holoneptf}
We can now use the results in appendix \ref{sec:HolographicOnePointFunctions} to extract the holographic one point functions\footnote{We take $\beta=r^{-1}[1+\tfrac{1}{4}|z_1|^2r^{-2}+\tfrac{1}{3}\operatorname{Re}(z_1 \bar{z_2}) r^{-3}+\dots]$
and $\s=x-\frac{1}{16}\partial_x |z_1|^2r^{-4}+\dots $ in order to go to the Fefferman-Graham coordinates used there.}. For the flat boundary metric
$ds^2_3=-dt^2+dy^2+d\sigma^2$, after writing
\begin{align}
z_1(\s)\equiv X_1(\s)+iY_1(\s)\,,
\end{align}
we find the stress tensor is given by
\begin{align}\label{eq:vevFmain}
\langle \mathcal{T}^{tt} \rangle = -\langle \mathcal{T}^{yy} \rangle= -2\partial_\s\left(X_1 Y_1\right)\,,\qquad
\langle \mathcal{T}^{\s \s} \rangle  = 4X_1\partial_\s{Y}_1\,.
\end{align}
The pseudo-scalar and scalar operator sources and VEVs are given by
\begin{align}
\mathcal{X}_s& =4\partial_\s{Y}_1 \,,\qquad
\mathcal{Y}_s  = Y_1\,,\nn
\langle \mathcal{O}_{\mathcal{X}} \rangle & = X_1\,,\qquad
\langle \mathcal{O}_{\mathcal{Y}} \rangle = 4\partial_\s{X}_1 \,.\label{eq:vevLmain}
\end{align}
The total conserved energy is given by
\begin{equation}
E_{\mathrm{susy}}  = 
\int \dd \s\, \dd y\, \langle \mathcal{T}^{tt}\rangle=-2\Delta y\int d\sigma\partial_\s\left(X_1 Y_1\right) \,,
\end{equation}
which demonstrates that the spatial average of the energy density over $\sigma$ vanishes for periodic solutions, or for 
those with sufficiently fast fall-off along the $\s$ direction. 

It is worth highlighting a feature of our construction which is implicit in the above analysis. 
For any given solution to the Helmholtz equation we can obtain
the holographic boundary, where $H=0$, and hence extract, in particular,  
the source $\mathcal{Y}_s$, and hence by supersymmetry, $\mathcal{X}_s$,  
associated to the spatially varying mass terms in ABJM theory. 
However, it is not easy to go in the other direction and obtain, for given supersymmetric source $\mathcal{Y}_s$ (and hence $\mathcal{X}_s$)
a solution to the Helmholtz equation which has those sources.

\subsection{Holographic lattices}
Assuming that $H$ is single valued\footnote{Much of the discussion here and in section \ref{sec5}
can be generalised to functions $H$ with a branch cut extending from a single delta function 
singularity out to infinity and with a finite number of sheets.} 
on the complex $w$ plane, any solution for which the locus $H=0$ is a closed curve 
will necessarily correspond to a holographic lattice solution, in which the spatial deformations in the dual theory are spatially periodic. Indeed
this immediately follows because a $2\pi$ rotation in the complex plane leaves the solution invariant\footnote{It would be interesting to explore other possible
classes of supersymmetric holographic lattice solutions. For example, one can consider solutions in which 
$H=0$ vanishes in the $w$-plane along a curve with a discrete $\mathbb{Z}$ isometry.}. 
The condition that the curve is closed, however, yields a non-trivial condition on the scalar sources and VEVs. Specifically, using \eqref{dwsz1} we have
\begin{equation}
\label{eq:ClosedCurveZConst}
0 = \oint dw = \int \frac{dw}{d\s} d\s = i \int z_1 d\s\,,
\end{equation}
and we conclude the zero Fourier-mode component of both sources and VEVs must vanish.

We now want to argue that any such holographic lattice, with $H=0$ on some closed curve on the plane,
must have a single isolated delta function source in the interior, just as in the Susy Q solution \eqref{susyq1},\eqref{susyq2}.
First suppose that there is no delta function source and $H>0$ throughout the domain, $D$, which is taken to be simply connected. If we multiply the Helmholtz equation with $H$ and then integrate over the domain we deduce
\begin{align}\label{posarg}
\int_D \nabla(H\nabla H)=\int_D [(\nabla H)^2+H^2] >0\,.
\end{align}
However, the left hand side is a boundary term, which vanishes since $H=0$ on the boundary. Thus, we conclude that such solutions without
delta function sources do not exist. 

Next suppose we have $n>0$ isolated delta function sources in the interior domain, $D$, with positive weight. We can also consider having $m$ poles with negative weight that are enclosed by circles with $H=0$. In other words $D$ has $m>0$ holes.
Recall that a necessary condition for $|z|<1$ is that $\nabla H\ne 0$, which is well defined everywhere 
in $D$, after excising small circles around each of the locations of the $n$ delta functions. 
Now around
the $H=0$ boundary, $\nabla H$ is an inward pointing normal and has winding number one. Furthermore, the winding number for both the $n$ excised circles and the $m$ interior holes where $H=0$ is also one.
Conservation of winding number as we contract the boundary curve then implies that we must have $n+m=1$.
We can eliminate the case $m=1$ and $n=0$ by using \eqref{posarg} and hence we conclude that we must have $m=0$ and
$n=1$ as claimed.

\subsection{Curves with corners}\label{cwcs}
When analysing specific solutions to the BPS equations we encounter situations in which 
the $H=0$ locus has corners. Working in polar coordinates, the zero locus associated with
the intersection of $n\ge 2$ curves where $H=0$ locally looks like $\rho^n\sin n\theta+\cO(\rho^{n+2})$. 
For $\rho \ll1$, the solution has the form
\begin{align}
ds^2 &\sim \frac{1}{\rho^{2n}\sin^2(n\theta)}\left[-dt^2+dy^2+\left(n^2\rho^{2n-2} \right)(d\rho^2 + \rho^2 d\theta^2)\right],\nn
~z &= -\frac{i}{n}
e^{-i(n-1)\theta}\rho\sin(n\theta).
\end{align}
We can consider a segment of the $AdS$ boundary to be located at $\theta=0$ and parametrised by $\rho$.
To make this more explicit we can change coordinates via 
\begin{align}
\theta = \frac{r}{n \s}+\ldots\,,
\qquad ~\rho = \s^{1/n}+\ldots\,,
\end{align}
with the boundary located at $r=0$ and parametrised by $\s$. In particular we can immediately obtain
the leading order falloff of the complex scalar field finding
\begin{align}\label{cornerz}
z_1(\s) = -\frac{i}{n}\s^{1/n-1}+\cO(\s^{3/n-1}).
\end{align}
Thus, we find for $n\ge2$ a singularity in the source and VEV, but it is integrable.

\section{Supersymmetric boomerang RG flows}\label{sec5}
It is now straightforward to construct infinite classes of boomerang RG flows, generalising the Susy Q solution. 
We consider holographic lattices for which $H$ vanishes along some closed curve in the complex plane, with $H$ single 
valued and $H>0$ in the interior. 
As we saw in the last section, this setup is associated with having spatially periodic  sources and VEVs in the dual field theory, 
which
have vanishing zero modes. Furthermore, the solutions necessarily have a single isolated delta function singularity in the interior and, 
just like we saw for
the Susy Q solution, this is associated with the same $AdS_4$ behaviour appearing in the IR as in the UV, and so 
we have a boomerang RG flow. For simplicity, we will take the delta function singularity to be located in the $w$ plane at $w=0$, which
can always be achieved after a translation of $w$.

Before illustrating with some simple explicit examples, we first obtain a general result concerning the refractive index of 
these boomerang flows. The refractive index in the direction of spatial modulation, $n_\s$, is defined to be the ratio of 
the speed of light in the direction of the spatial modulation, $\sigma$, in the UV and the IR, as determined by the 
corresponding $AdS_4$ regions of the solution. 
To extract an expression for $n_\s$ in terms of $H$ we proceed as follows. 
We first consider the $AdS_4$ UV boundary metric \eqref{expnigbet}. 
We define $\Delta\s$ to be the proper length with respect to the three-dimensional
boundary metric, located at $H=0$, of the direction parametrised by the spatial coordinate $\s$. Using \eqref{sigdef} we have
\begin{equation}
\Delta \s \equiv  \oint_{H=0} d\s = 2 \oint_{H=0} |\partial_w H| \left| \frac{dw}{dl} \right| dl = 2 \oint_{H=0} |\partial_w H| d\tilde{s}\,,
\end{equation} 
where $l$ is an arbitrary length parameter, while $\tilde s$ is the length parameter in the (flat) $w$-plane. 
We next introduce the periodic coordinate $\theta\equiv 2\pi \frac{\s}{\Delta\sigma}$, with period $2\pi$. The
UV $AdS_4$ boundary metric can then be written, as $\n\to 0$, in the form
\begin{equation}
ds^2\sim \frac{1}{\n^2} \left[ -dt^2 + dy^2 + \left( \frac{\Delta\sigma}{2\pi} \right)^2 d\theta^2 + d\n^2  \right]\,.
\end{equation}

We now want to compare this with the $AdS_4$ factor in the IR. 
The log singularity is assumed
to be located at the origin in the $w$ plane, as already mentioned, and so we write
$H \sim - \cn \log |w|$ and $\cn>0$, as $w\to 0$. Switching to the 
coordinates given by $ w \equiv e^{-\frac{1}{\cn\rho} + i \theta}$, with $\theta$ again having period $2\pi$, we can write 
the metric and scalar field $z$, as $w\to 0$, in the form
\begin{align}
\label{eq:IRMetricOnWPlane}
ds^2 &\sim \frac{d\rho^2}{\rho^2} + \rho^2 (-dt^2 + dy^2 + h_0^2 d\theta^2)\,, \nn
z &\sim \frac{1}{\cn\rho} e^{-\frac{1}{\cn\rho}+i\theta}\,.
\end{align}
Hence, by definition, the refractive index of the boomerang RG flow is given by
\begin{equation}\label{refingen}
n_\s
= \frac{2\pi \cn}{\Delta\s} = 
\frac{\pi \cn}{\oint_{H=0} |\partial_w H| d\tilde{s}}\,.
\end{equation}

We can now obtain an analytic expression for $n_\sigma$ for the Susy Q solution, which was found
numerically in \cite{Gauntlett:2018vhk}. Starting with \eqref{susyq1} we immediately find that $\sigma=\theta$. Furthermore,
$z_1(\s)=ce^{i\s}$, which allows us to obtain the sources and VEVs. Finally, 
\eqref{refingen} gives the simple result $n_\s=I_0(c)$, where $c$ is the parameter appearing in \eqref{susyq1}.

We can also show in general that $ n_\s \geq 1$, as seen in the Susy Q solution \cite{Gauntlett:2018vhk},
as well as for other non supersymmetric boomerang RG flows \cite{Donos:2017ljs,Donos:2017sba}. 
To see this we let $l$ be a counter-clockwise parameter on the $H=0$ curve in the complex plane. Then, using the fact that
$\operatorname{Re}\left(\partial_w H \frac{dw}{dl}\right) = 0$ (from \eqref{sigdef})
and $\operatorname{Im}\left(\partial_w H \frac{dw}{dl}\right) \leq 0$ (since $H$ is necessarily decreasing as one moves from
the interior to the exterior of the $H=0$ curve), we
can write 
\begin{equation}
\Delta\s = \oint_{H=0} \left[ i \partial_wH dw - i \partial_{\bar{w}} H d\bar{w} \right].
\end{equation}
After using Stokes theorem and the Helmholtz equation we have
\begin{align}
\Delta\s&= -\int_{H>0} \partial_w \partial_{\bar{w}} H\, 2i  dwd\bar{w}\,,\nn
&= 
- \int_{H>0} \left(H - 2\pi \cn\, \delta(w,\bar{w}) \right)\frac{i}{2}dwd\bar{w} \leq 2 \pi \cn \,,
\end{align} 
and hence $n_\s \geq 1$.

In fact, a closed analytic expression can be derived for $n_\sigma$ just in terms of the source $Y_1(\sigma)$ alone (or alternatively just in terms of $X_1(\sigma)$, which determines the VEVs). To see this, we first note that solutions to 
the Helmholtz equation \eqref{helm} are associated with an infinite family of closed one-forms given by
\begin{equation}
M_{(\alpha)} \equiv e^{-\frac{i}{2}\alpha w + \frac{i}{2\alpha}\bar{w}} \Big( \partial_w H dw + \frac{i}{2\alpha} H d\bar{w} \Big)\,,
\end{equation}
which we have labeled by the arbitrary parameter $\alpha \in \mathbb{C}$. Indeed taking the exterior derivative of these one-forms 
we find
\begin{equation}
dM_{(\alpha)} = e^{-\frac{i}{2}\alpha w + \frac{i}{2\alpha}\bar{w}} \Big( \frac{1}{4}H - \partial_{w\bar{w}}H \Big) dw \wedge d\bar{w}\,,
\end{equation}
and hence $dM_{(\alpha)}=0$ if and only if the Helmholtz equation is satisfied\footnote{In fact, such a family of one-forms exists for any linear PDE in two dimensions with constant coefficients \cite{2002FokasZyskin}.}. Integrating $M_{(\alpha)}$ on the $H=0$ boundary curve and using \eqref{vec1}
we get
\begin{equation}
\oint_{H=0} M_{(\alpha)} = -\frac{i}{2} \oint_{H=0} e^{-\frac{i}{2}\alpha w(\sigma) + \frac{i}{2\alpha}\bar{w}(\sigma)} d\sigma\,.
\end{equation}
Using Stokes theorem, the fact that the Helmholtz equation has a delta function source at $w=0$,
as well as \eqref{refingen}, then results in an infinite set of constraints on $w(\sigma)$ given by
\begin{equation}\label{sourcevevconstraints}
\frac{1}{\Delta\sigma} \oint_{H=0} e^{-\frac{i}{2}\alpha w(\sigma) + \frac{i}{2\alpha}\bar{w}(\sigma)} d\sigma = n_\sigma \,,
\end{equation}
for arbitary $\alpha \in \mathbb{C}$. In particular, taking $\alpha = \pm i$ we obtain:
\begin{equation}
n_\sigma = \frac{1}{\Delta\sigma}\oint_{H=0} e^{\pm\operatorname{Re}w(\sigma)} d\sigma \,,
\end{equation}
from which we can easily conclude that $n_\sigma\ge 1$ as already shown above.
Also notice, that the arbitrary constant in the integration of \eqref{dwsz1} can be fixed by imposing $\int e^{\operatorname{Re}(w(\sigma))} d\sigma = \int e^{-\operatorname{Re}(w(\sigma))} d\sigma $. We can now use these results to obtain the following expression
for $n_\sigma$ in terms of the source function $Y_1(\sigma)$ alone
\begin{align}
n_\sigma&=\frac{1}{\Delta\sigma}\Big[\oint_{H=0}d\sigma e^{\int_0^\sigma Y_1(\sigma')d\sigma'}\Big]^{1/2}
\Big[\oint_{H=0}d\sigma e^{-\int_0^\sigma Y_1(\sigma')d\sigma'}\Big]^{1/2}\,,\nn
&=
\frac{1}{\Delta\sigma}\Big[\oint_{H=0}d\sigma d\tilde \sigma e^{\int_{\tilde\sigma}^\sigma Y_1(\sigma')d\sigma'}\Big]^{1/2}\,.
\end{align}
For example, substituting $Y_1(\sigma)=c \cos\sigma$, as in the Susy Q solution, we immediately obtain 
$n_\s=I_0(c)$ as before.
A similar formula involving $X_1(\sigma)$, and hence the VEVs, can be derived by choosing $\alpha=\pm 1$.

\subsection{Deformations of Susy Q}

Suppose we solve the Helmholtz equation using polar coordinates on the complex plane with $w=\rho e^{i\theta}$. 
By separation of variables we obtain solutions in terms of modified Bessel functions
of the form $I_n(\rho)e^{in\theta}$, $K_n(\rho)e^{in\theta}$ for $n\ge 0$.
Now as we approach the origin, $\rho\to 0$, we have $I_n(\rho)\to 0$, but $K_n(\rho)\to \infty$. We want to keep 
one term of the form $K_0(\rho)$ to ensure we have a single isolated delta function singularity. We can discard the
terms involving $K_n(\rho)$, for $n>0$ since they are associated with regions surrounding the divergence with $H<0$.
We therefore consider
\begin{align}\label{helmexp}
H=d_0 K_{0}(\rho) +\sum_{n=0}^\infty I_n(\rho)(a_n\cos n\theta +b_n \sin n\theta)\,,
\end{align}
with $d_0>0$.
A particular example is the susy Q solution with the only non-vanishing term in the sum arising from $a_0\ne 0$. 

Although this will certainly solve the BPS equations, it is not a simple matter to determine the values of $d_0,a_n,b_n$ for which $H=0$ actually vanishes on a given closed curve. Having done that, one then needs to carry out the analysis in section
\ref{gencoms} in order to extract out the sources and VEVs for the dual operators in ABJM theory.  
Progress can be made by considering small perturbations about the Susy Q solution but we will not present any details here\footnote{As an alternative, one may try to solve the infinite set of constraints for the sources and VEVs given in equation \eqref{sourcevevconstraints} for an arbitrary deformation of the Susy Q solution.}.

A general point that can be made is the following. For each mode $n$ in \eqref{helmexp} we have two pieces of real data $a_n$, $b_n$ to specify. On the other hand, the sources and the VEVs of the two operators in the dual theory are given by $z_1(\s)$, $z_2(\s)$ and are associated with eight pieces of real data for each Fourier mode. However, since the BPS equations imply that $z_2=  i\frac{d z_1}{d\s}$ we have just four pieces of independent data.
The restriction to a supersymmetric boomerang RG flow therefore fixes two of the four boundary pieces of data for each mode, rather than just placing a restriction on the zero modes. Generically, then, we deduce that for given periodic sources, with vanishing zero modes, we must tune the VEVs to be specific 
periodic functions with vanishing zero modes in order to get a supersymmetric boomerang RG flow. Thus, the possible naive intuition
that there could be many choices of VEVs for the given sources that 
give rise to a supersymmetric boomerang RG flow is not correct.

\begin{figure}[h!]
\centering
\raisebox{-0.5\height}{\includegraphics[scale=0.35]{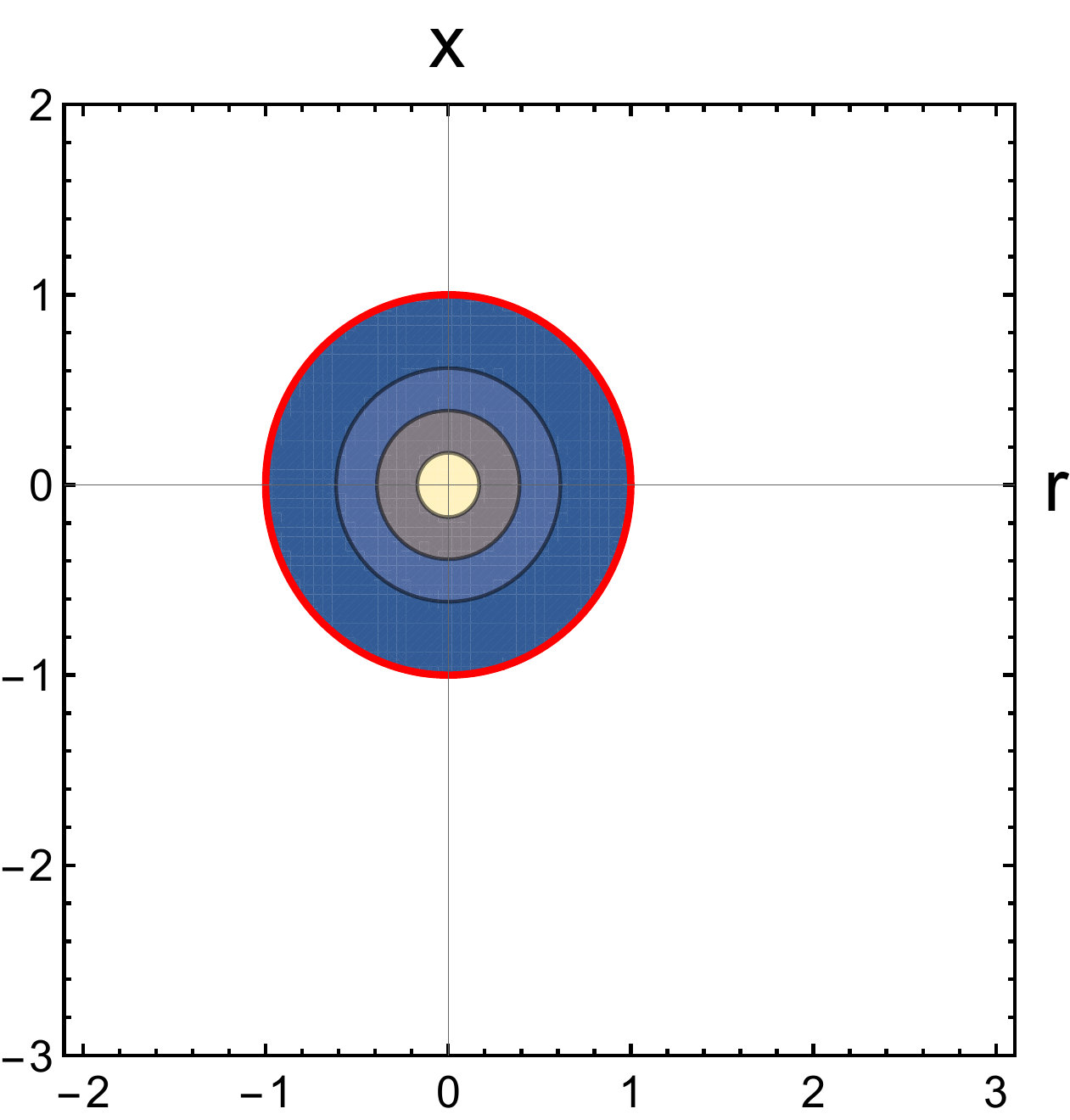}\qquad}
\raisebox{-0.5\height}{\includegraphics[scale=0.45]{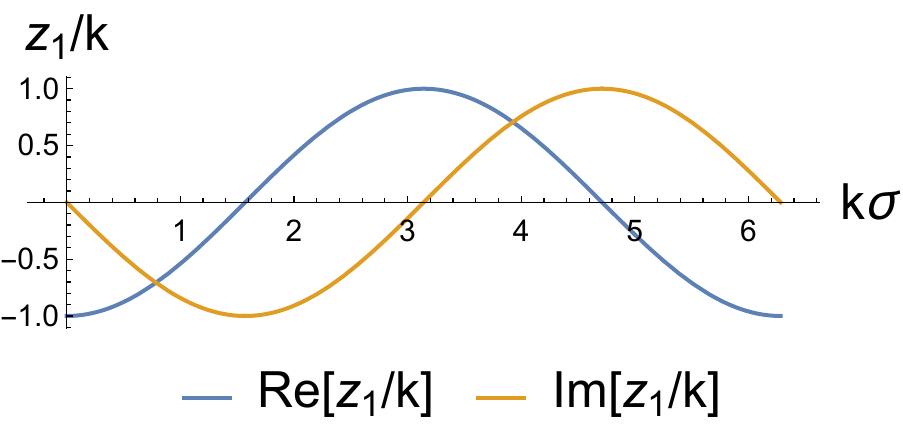}}\\
\raisebox{-0.5\height}{\includegraphics[scale=0.35]{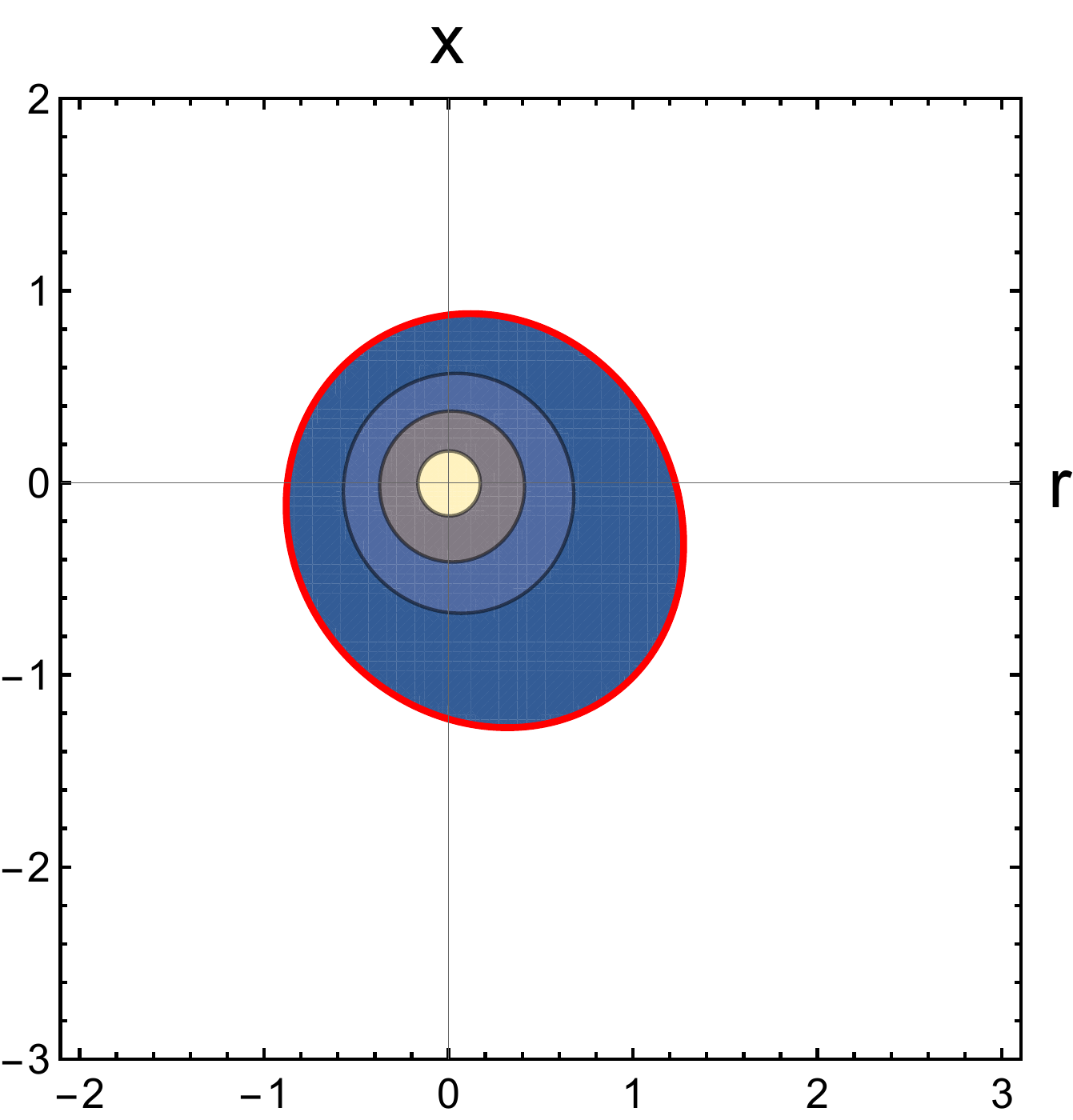}}\qquad
\raisebox{-0.5\height}{\includegraphics[scale=0.45]{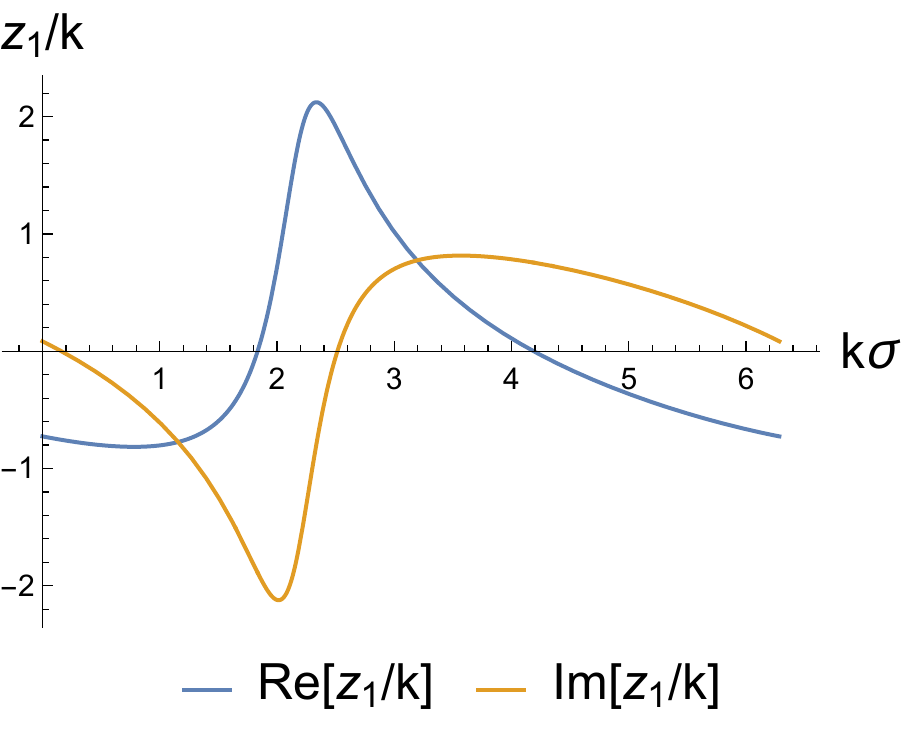}}\\
\raisebox{-0.5\height}{\includegraphics[scale=0.35]{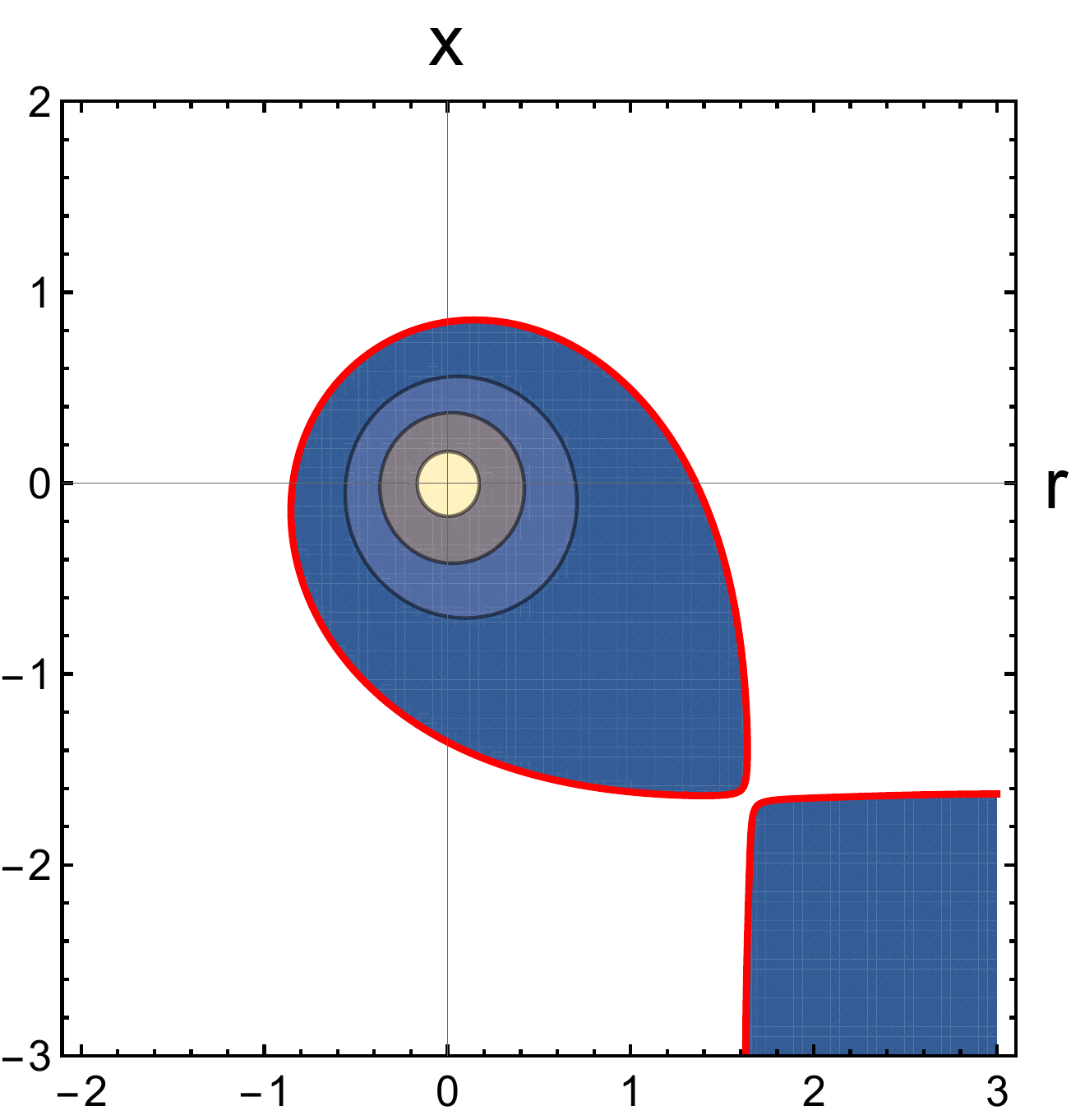}}\qquad
\raisebox{-0.5\height}{\includegraphics[scale=0.45]{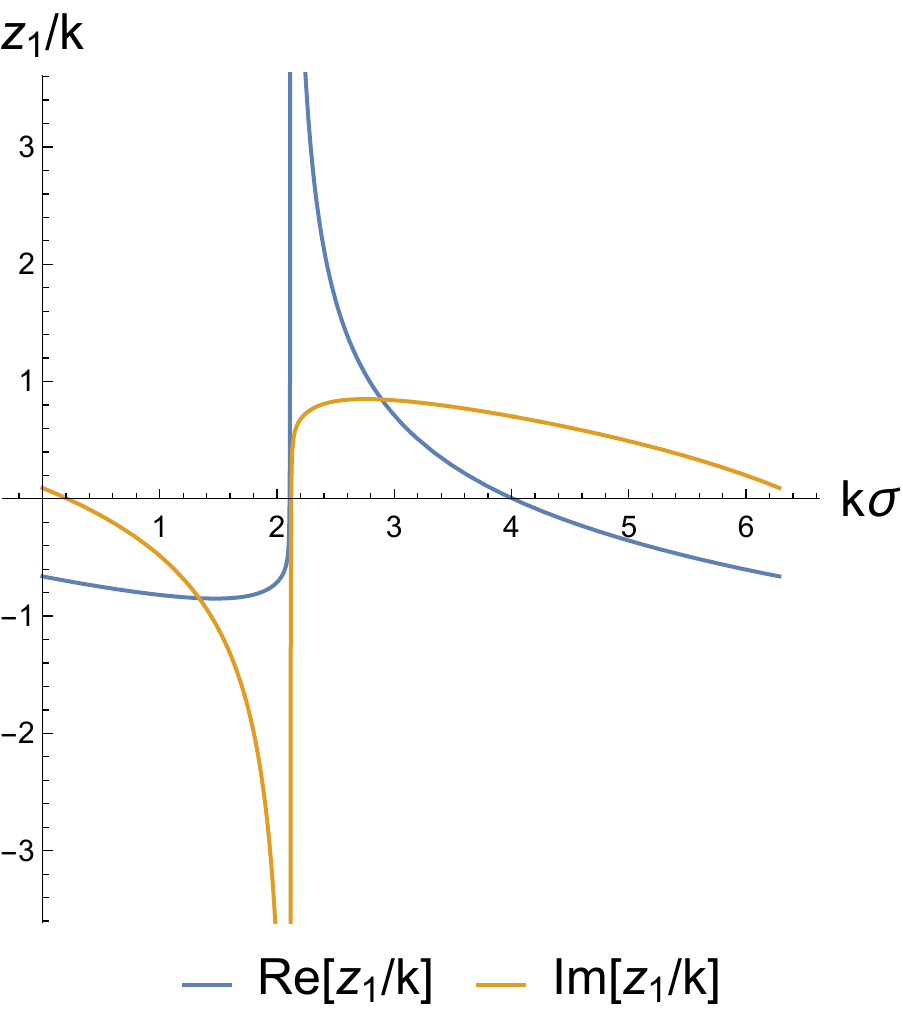}}\\
\caption{\label{fig:one}Boomerang RG flow solutions for deformations of the Susy Q solution given in \eqref{defsusyq} with $c=1$ and varying $b_1$. On the left are contour plots of $H$ in the complex $w$-plane, with the $H=0$ locus marked with a red line.
The right hand plots show the
behaviour of $z_1(\s)=X_1(\s)+iY_1(\s)$, which fixes the one point functions as discussed in section \ref{holoneptf}, and $k\equiv 2\pi/\Delta\s$. In particular, we have
$\mathcal{Y}_s  = Y_1$ and $\langle \mathcal{O}_{\mathcal{X}} \rangle = X_1$.
From top to bottom, the plots are for $b_1=0$ (i.e. Susy Q), $b_1=0.4$ and $b_1=0.52$. For larger values of $b_1$ the $H=0$ locus
is no longer a closed curve, the solution is no longer associated with a spatially periodic deformation and singularities appear.}
\end{figure}

We will now analyse some specific examples of boomerang RG flows and extract out the physical quantities in the 
dual theory using numerics. We consider a deformation of a specific Susy Q solution with the addition of just a single Fourier mode
\begin{align}\label{defsusyq}
H=H_\text{Susy Q}(c)+ b_1I_1(\rho)  \cos (\theta +\frac{\pi}{4})\,.
\end{align}
To illustrate we set $c=1$. We then find that we can vary $b_1$ from zero up to a maximum value $b_1\sim 0.5202$, at which point
the closed curve where $H=0$ meets another $H=0$ locus, as shown in figure \ref{fig:one}. 
The phase shift of $\pi/4$ has been chosen simply so that the intersecting lines are parallel with the $r$ and $x$ axes.
For larger values of $b_1$ the $H=0$ locus is no longer a
closed curve and hence is no longer associated with a spatially periodic deformation. One can check that $|z|<1$ up until the maximum 
value of $b_1$, but this is no longer the case for larger values 
and the solutions develop singularities.
Figure \ref{fig:one} also shows the behaviour of $z_1(\sigma$), which
fixes the one point functions as discussed in section \ref{holoneptf}. As the maximum value of $b_1$ is reached we see that the source and the VEV develop a square root
singularity, as anticipated from \eqref{cornerz}.

\section{Discussion}\label{sec6}
In this paper we have constructed rich classes of new supersymmetric solutions of $D=11$ supergravity
that are holographically dual to ABJM theory which has been deformed by spatially
dependent masses. Remarkably, the entire construction reduces to finding solutions to the Helmholtz equation on the plane.
We have discussed infinite new classes of supersymmetric boomerang RG flows, generalising the
Susy Q solution \cite{Gauntlett:2018vhk}. In these solutions both the sources and the VEVs in the dual ABJM theory are spatially periodic
and with vanishing zero mode. In addition the energy density averaged over a spatial period vanishes exactly
and we also showed the refractive index is always bigger than 1.

It would be very interesting to further investigate our solutions in the weakly coupled ABJM theory.
To start with one should be able to derive the relations for the VEVs for the one point functions 
that we derived from the gravity perspective given in \eqref{eq:vevFmain},
\eqref{eq:vevLmain}. 
One can show that the deformed BPS equations imply, in particular, that
\begin{align}
\partial_x \mathrm{Tr}\big( M_A{}^B Y^A Y^\dagger_B \big) = \frac{8 \pi}{q} \mathrm{Tr} \big(M_A{}^B Y^C Y^\dagger_{[C} Y^A Y^\dagger_{B]}\big)+2 m \mathrm{Tr} (Y^A Y^\dagger_A)\,,
\end{align}
which is consistent with the last line of \eqref{eq:vevLmain}. Indeed, recalling \eqref{eq:150}, on the left hand side is the spatial derivative of the operator $\mathcal{O}_\mathcal{X}^{\Delta = 1}$ (up to normalisation), while on the right hand side is a linear combination of the bosonic part of the operator $\mathcal{O}_\mathcal{Y}^{\Delta = 2}$ together with the unprotected operator $|Y|^2$, which is not visible in the supergravity limit. More ambitiously, one can try to extend the analysis of the vacuum structure
for spatially independent mass deformations studied in \cite{Gomis:2008vc,Kim:2010mr} to the spatially dependent case.

The boomerang RG flows of this paper, as well as other holographic constructions \cite{Chesler:2013qla,Donos:2014gya,Donos:2016zpf,Donos:2017sba,Gauntlett:2018vhk}, are associated with gauge theories in a large $N$ limit. However, 
analogous results for certain condensed matter systems with periodic 
disorder are also known. For example, in the context of systems with
low energy excitations described by a $d=2+1$ Dirac fermion, it was shown that random spatially dependent mass terms which
average to zero are marginally irrelevant perturbations \cite{PhysRevB.50.7526}. For the case of $m(x,y) = m_0 \cos(k x)$ it was observed that band repulsion from the mass term suppresses the spectral density at low energies \cite{2013VafekVishwanath}. This can be contrasted with the case of adding a spatially periodic chemical potential of the from $\mu(x,y) =V \cos(k x)$ 
which causes an increase of the density of states at low energy \cite{2009BreyFertig}. In fact for this specific chemical 
potential deformation it was found that for large enough $V/k$ new Dirac cones appear at nonzero momentum in the $y$ direction
\cite{2009BreyFertig,PhysRevB.81.075438}. It would be interesting to further study the known holographic solutions, in this context, 
to see whether they share any universal features with solid state systems. The new solutions we have presented in this paper
are particularly interesting  in this regard since, quite remarkably, the ground states of the spatially deformed systems
are described by analytic solutions of $D=11$ supergravity.

The new supersymmetric solutions were constructed using the $SO(4)\times SO(4)$ invariant sector of $\mathcal{N}=8$
gauged supergravity. It may be possible to generalise our construction in several different directions.
One possibility is to consider more general $SO(4)\times SO(4)$ invariant configurations in $D=11$ supergravity.
Another possibility is to investigate similar constructions within other sectors of 
$\mathcal{N}=8$ gauged supergravity. For example, supersymmetric Janus solutions have been
constructed in the $SU(3)\times U(1)^2$ and $G_2$ invariant sectors which preserve four and two supersymmetries,
respectively \cite{Bobev:2013yra}. On the other hand we should note that these sectors do not admit a simple Susy Q solution \cite{Gauntlett:2018vhk}, and so there
may be some significant differences for these cases compared 
to the $SO(4)\times SO(4)$ invariant sector.

\subsection*{Acknowledgments}
We would like to thank Elias Kiritsis and Daniel Waldram for helpful discussions.
IA, JPG and MR  are supported by the European Research Council under the European Union's Seventh Framework Programme (FP7/2007-2013), ERC Grant agreement ADG 339140. JPG is also supported by STFC grant ST/P000762/1, EPSRC grant EP/K034456/1, as a KIAS Scholar and as a Visiting Fellow at the Perimeter Institute.  JPG acknowledges hospitality from the KITP and
support from National Science Foundation under Grant No. NSF PHY-1748958. CR
thanks the Crete Center for Theoretical Physics for hospitality.

\appendix

\section{Deriving the BPS equations}\label{appa}
In this appendix we derive the BPS equations and determine the amount of supersymmetry that is preserved.
We first carry out the analysis within $\mathcal{N}=1$ gauged supergravity in $D=4$ before discussing
$\mathcal{N}=8$ gauged supergravity.

\subsection{$\mathcal{N}=1$ $D=4$ supergravity conventions}
We begin with a few general formulae, mostly using the conventions given in \cite{fvpbook}.
We consider $\mathcal{N}=1$ supergravity in $D=4$ coupled to an arbitrary number of chiral multiplets with the bosonic part of the
action given by 
\begin{equation}\label{eq:Sgfirstone}
S= \int  \dd^4 x \sqrt{-g}\Big(R - G_{\alpha\bb}\partial_\mu z^\alpha \partial^\mu \bar{z}^{\bb}-\mathcal{V}\Big) \,.
\end{equation}
The complex scalar fields $z^\alpha$ parametrise a K\"ahler manifold with metric given by
\begin{equation}\label{eq:Nis1data}
G_{\alpha\bb} = 2\partial_\alpha\partial_{\bb}\mathcal{K}\,,
\end{equation}
where $\mathcal{K}$ is the K\"ahler potential\footnote{Note that we have set $16\pi G=2\kappa^2=1$ and
our K\"ahler potential, $\mathcal{K}$, is related to the one in \cite{fvpbook}, $\mathcal{K}_{there}$, via
$\mathcal{K}=\kappa^2 \mathcal{K}_{there}=(1/2)\mathcal{K}_{there}$.}. The potential $\mathcal{V}$ is given by
\begin{equation}\label{potterms}
\mathcal{V} = 4 G^{\alpha\bb}\partial_\alpha\mathcal{W}\partial_{\bb}\mathcal{W}-\frac{3}{2}\mathcal{W}^2\,,\qquad
\mathcal{W} =-e^{\mathcal{K}/2}|W|\,,
\end{equation}
where $W$ is the superpotential which is a holomorphic function of the $z^\alpha$.

We use gamma matrices given by
\begin{equation}\label{gammmat}
\Gamma^{\hat \mu} = 
\begin{pmatrix}
0 & \sigma^{\hat \mu} \\
\bar{\sigma}^{\hat \mu} & 0
\end{pmatrix}\,,
\qquad \mathrm{with}\qquad \sigma^{\hat \mu} = (1,\vec{\sigma}), \qquad \bar{\sigma}^{\hat \mu} = (-1, \vec{\sigma})\,,
\end{equation}
where the hatted indices refers to an orthonormal frame,
and $\vec{\sigma}$ are the three Pauli matrices. The four-component Majorana spinor, parametrising the supersymmetry transformations, can be expressed in terms of chiral components via
\begin{equation}\label{majcon}
\hat{\epsilon} = (\epsilon,\tilde{\epsilon})^T
\qquad \mathrm{with} \qquad \tilde{\epsilon} = i\sigma_{2} \,\epsilon^*\,,
\end{equation}
and the chiral projections are given by
\begin{equation}
\Gamma^5 = -i \Gamma^{\hat 0}\Gamma^{\hat 1}\Gamma^{\hat 2}\Gamma^{\hat 3}, \qquad P_{L/R} = \frac{1}{2}\left(1\pm\Gamma^5 \right)\,.
\end{equation}
The variations for the fermions as chiral spinors can then be written as
\begin{align}\label{Susyvars}
\delta \psi_\mu & =\, \left(\nabla_\mu -\frac{3}{2}i\mathcal{A}_\mu \right) \epsilon + \frac{1}{4}\sigma_\mu e^{\mathcal{K}/2} W \tilde{\epsilon}\,,\nn
\delta\tilde{ \psi}_\mu & =\, \left(\tilde{\nabla}_\mu +\frac{3}{2}i\mathcal{A}_\mu \right) \tilde{\epsilon} + \frac{1}{4}\bar{\sigma}_\mu e^{\mathcal{K}/2} \bar{W} \epsilon\,,\nn
\sqrt{2}\delta\chi^\alpha & = \,\sigma^\mu\partial_\mu z^\alpha\tilde{\epsilon} -e^{\mathcal{K}/2}G^{\alpha\bb}D_{\bb}\bar{W}\epsilon \,,\nn
\sqrt{2}\delta\tilde{\chi}^{\bb} & = \,\bar{\sigma}^\mu\partial_\mu \bar{z}^{\bb}\epsilon - e^{\mathcal{K}/2}G^{\alpha\bb}D_{\alpha}W\tilde{\epsilon}\,,
\end{align}
where the various covariant derivatives are defined as
\begin{equation}
\nabla_\mu = \partial_\mu +\frac{1}{4}\omega_\mu\,^{\hat\nu\hat\rho}\sigma_{[\hat\nu}\bar{\sigma}_{\hat\rho]}, \qquad \tilde{\nabla}_\mu = \partial_\mu +\frac{1}{4}\omega_\mu\,^{\hat\nu\hat\rho}\bar{\sigma}_{[\hat\nu}\sigma_{\hat \rho]}, \qquad D_\alpha = \partial_\alpha +\partial_\alpha\mathcal{K},
\end{equation}
and the K\"ahler connection is given by
\begin{equation}
\mathcal{A}_\mu = \frac{i}{6}\sum_\alpha (\partial_\alpha\mathcal{K}\partial_\mu z^\alpha
-\partial_{\bar \alpha}\mathcal{K}\partial_\mu \bar z^{\bar \alpha})\,.
\end{equation}

\subsection{Derivation of the BPS equations in $\mathcal{N}=1$ supergravity}
In principle the results of this section can be obtained from
the general analysis of supersymmetric configurations of $\mathcal{N}=1$ supergravity that was carried out in \cite{Gran:2008vx}, however, we have carried out the analysis {\it ab initio}.

The $D=4$ theory of interest, with bosonic action as in \eqref{eq:Sg2bulk}, can be recast in the form
given in \eqref{eq:Sgfirstone} by taking 
\begin{align}\label{kanddubya}
e^{\mathcal{-K}/2}=2^{3/2}(1-|z|^2)^{1/2}\,,\qquad W=2^{5/2}\,.
\end{align}
In this appendix, we will analyse the conditions for supersymmetry for general $\mathcal{K}$ and 
holomorphic $W$, before
restricting to \eqref{kanddubya}.
We consider the ansatz 
\begin{align}\label{Susyqansatzapp}
ds^2&=e^{2A}(-dt^2+dy^2)+e^{2V} dx^2 +N^2dr^2\,,
\end{align}
with $A$, $N$, $V$ and the scalar $z$ all functions of $r,x$.
We use the orthonormal frame
$(e^{\hat{t}},e^{\hat{x}},e^{\hat{y}},e^{\hat{r}})= 
(e^A dt, e^V dx, e^A dy, Ndr)$ and, consistent with \eqref{gammmat},
$(\sigma^{\hat x},\sigma^{\hat y},\sigma^{\hat r})=\vec{\sigma}$.
We begin by imposing the projection
\begin{align}\label{basicproj}
\Gamma^{\hat{t}\hat{y}}\hat\epsilon=-\hat\epsilon\,,
\end{align}
which implies that on the chiral spinors we have $\sigma^{\hat{y}}\epsilon=-\epsilon$ and $\sigma^{\hat{y}}\tilde \epsilon=\tilde\epsilon$. We also assume that the Killing spinor is independent of $t,y$. 

After substituting into the supersymmetry transformations \eqref{Susyvars}
we get the following BPS equations (plus the complex conjugates). 
From the
$t,y$ components of the variation of the gravitino we get 
\begin{align}\label{tyeq}
\frac{1}{2}e^{\mathcal{K}/2}W\sigma^{\hat r}\tilde\epsilon=\left(ie^{-V}\partial_x A-N^{-1}\partial_r A\right)\epsilon\,,
\end{align}
while from the $r$ and $x$ components we get 
\begin{align}\label{req}
N^{-1}\partial_r\epsilon-\frac{i}{2}e^{-V}N^{-1}\partial_x N\epsilon-\frac{3i}{2}\mathcal{A}_{\hat r}\epsilon+
\frac{1}{4}e^{\mathcal{K}/2}W\tilde\epsilon&=0\,,\nn
-ie^{-V}\partial_x\epsilon+\frac{1}{2}N^{-1}\partial_r V\epsilon-\frac{3}{2}\mathcal{A}_{\hat x}\epsilon+\frac{1}{4}e^{\mathcal{K}/2}W\sigma^{\hat r}\tilde\epsilon&=0\,,
\end{align}
where
\begin{equation}
\mathcal{A}_{\hat r} = \frac{i}{6}N^{-1}(\partial \mathcal{K}\partial_r z
-\bar\partial\mathcal{K}\partial_r \bar z)\,,
\qquad
\mathcal{A}_{\hat x} = \frac{i}{6}e^{-V}(\partial \mathcal{K}\partial_x z
-\bar\partial\mathcal{K}\partial_x \bar z)\,.
\end{equation}
Finally, from the variation of the dilatino we get 
\begin{align}\label{sceq}
\left(ie^{-V}\partial_x z+N^{-1}\partial_r z\right)\sigma^{\hat r}\tilde\epsilon
-e^{\mathcal{K}/2}G^{-1}\bar D \bar W\epsilon=0\,.
\end{align}
Notice that if we use \eqref{tyeq} in \eqref{req} and \eqref{sceq} then the latter just depend on $\epsilon$.

Recalling the Majorana condition \eqref{majcon}, $\tilde \epsilon = i \sigma^{\hat y}\epsilon^*$,
from \eqref{tyeq} we deduce that 
\begin{align}
\left|
\left(\tfrac{1}{2}e^{\mathcal{K}/2}W\right)^{-1}\left(ie^{-V}\partial_x A-N^{-1}\partial_r A\right)\right|^2=1\,.
\end{align}
This can be solved by imposing an additional restriction on the chiral spinors
\begin{align}
\sigma^{\hat r}\tilde \epsilon=-e^{i\xi}\epsilon\,,
\end{align}
with $\xi=\xi(r,x)$ satisfying 
\begin{align}
e^{-V}\partial_xA=-\frac{1}{2}e^{\mathcal{K}/2}\text{Im}(We^{i\xi})\,,
\qquad
N^{-1}\partial_r A=\frac{1}{2}e^{\mathcal{K}/2}\text{Re}(We^{i\xi})\,.
\end{align}
Note that this implies the integrability condition
\begin{align}\label{intcd}
\partial_r \left[e^{\mathcal{K}/2}{e^{V}\text{Im}(We^{i\xi})} \right] 
+\partial_x \left[e^{\mathcal{K}/2}N\text{Re}(We^{i\xi})  \right] =0\,.
\end{align}
Using \eqref{basicproj}, we also have $\epsilon^*=-e^{i\xi}\sigma^{\hat x}\epsilon$.

Continuing, \eqref{sceq} now takes the form
\begin{align}\label{sceq2}
\left(ie^{-V}\partial_x z+N^{-1}\partial_r z\right)e^{i\xi}+e^{\mathcal{K}/2}G^{-1}\bar D\bar W=0\,.
\end{align}
With some foresight we can rewrite \eqref{req} in the form
\begin{align}
\label{jhgn2}
e^{-i\xi/2}e^{A/2}\partial_r&(e^{-A/2}e^{i\xi/2}\epsilon)
-\frac{iN}{2}\left[N^{-1}\partial_r\xi+e^{-V}N^{-1}\partial_x N
+3\mathcal{A}_{\hat r}
+\frac{1}{2}e^{\mathcal{K}/2}\text{Im}(We^{i\xi})
\right]\epsilon=0\,,\nn
e^{-i\xi/2}e^{A/2}\partial_x&(e^{-A/2}e^{i\xi/2}\epsilon)
-\frac{ie^V}{2}\left[e^{-V}\partial_x\xi-N^{-1}\partial_r V
+3\mathcal{A}_{\hat x}
+\frac{1}{2}e^{\mathcal{K}/2}\text{Re}(We^{i\xi})\right]\epsilon=0\,.
\end{align}
By taking the complex conjugates of these two equations, and using $\epsilon^*=-e^{i\xi}\sigma^{\hat x}\epsilon$, we deduce that in each expression, each of the two terms must separately vanish.
Thus, the Killing spinor must take the form
\begin{align}\label{fundepks}
\epsilon=e^{A/2}e^{-i\xi/2}\eta_0,\qquad
\tilde \epsilon=-e^{i\xi}\sigma^{\hat r}\epsilon\,,
\end{align}
where $\eta_0$ is a constant chiral spinor with $\sigma^{\hat{y}}\eta_0=-\eta_0$ and $\eta_0^*=-\sigma^{\hat x}\eta_0$.
Moreover, the combined system of BPS equations is then given
\begin{align}\label{bpsderxi1appa}
\left(N^{-1}\partial_r A-ie^{-V}\partial_x A\right)=\frac{1}{2}e^{\mathcal{K}/2}We^{i\xi}\,,\nn
\left(ie^{-V}\partial_x z+N^{-1}\partial_r z\right)e^{i\xi}+e^{\mathcal{K}/2}G^{-1}\bar D\bar W&=0\,,\nn
N^{-1}\partial_r\xi+e^{-V}N^{-1}\partial_x N
+3\mathcal{A}_{\hat r}
+\frac{1}{2}e^{\mathcal{K}/2}\text{Im}(We^{i\xi})&=0\,,\nn
e^{-V}\partial_x\xi-N^{-1}\partial_r V
+3\mathcal{A}_{\hat x}
+\frac{1}{2}e^{\mathcal{K}/2}\text{Re}(We^{i\xi})&=0\,.
\end{align}
These equations are not all independent.
This can be seen using \eqref{intcd} and also the fact that
\begin{align}
\partial_r\mathcal{K}+i\partial_x\mathcal{K}+6e^{V}\mathcal{A}_{\hat x}-6iN\mathcal{A}_{\hat r}
=2\partial\mathcal{K}(\partial_r z+i\partial_x z)\,.
\end{align}

We also note that the conditions on the chiral spinors given above can be recast in terms of four component Majorana spinors, $\hat\epsilon$, as
\begin{align}
\Gamma^{\hat{t}\hat{y}}\hat\epsilon=-\hat\epsilon\,,\qquad
{}[-\sin\xi\Gamma^{\hat x}+\cos\xi\Gamma^{\hat r}]\hat\epsilon=-\hat\epsilon\,.
\end{align}

If we substitute \eqref{kanddubya} into \eqref{bpsderxi1appa}, then we obtain the BPS equations given in
\eqref{bpsderxi1}. We also note that if, as in the text, we consider the 
gauge $N=e^V$ then the BPS equations can be cast 
into the form
\begin{align}\label{susygenintapp}
\partial A&=B\,,\nn
\bar \partial z&=-2G^{-1}\frac{\bar D\bar W}{\bar W} \bar B\,,\nn
\bar \partial B&=-\left(1-2G^{-1}\frac{|D W|^2}{|W|^2}\right) B\wedge \bar B\,,
\end{align}
where the $(1,0)$ form $B$ is defined as
\begin{align}\label{Bdefapp}
B\equiv \frac{1}{4}e^{V+\mathcal{K}/2+i\xi}W dw\,.
\end{align}
If we substitute \eqref{kanddubya} into \eqref{susygenintapp}, then we obtain the BPS equations given in
\eqref{susygenint}.

\subsection{Supersymmetry in $\mathcal{N}=8$ gauged supergravity}\label{appb}
Here we discuss the preserved supersymmetry from the perspective of the $\mathcal{N}=8$ theory, referring
to appendix B of \cite{Gauntlett:2018vhk}, for further details.
The supersymmetry variations of the fermions are written in terms of eight left/right chiral spinor parameters $\epsilon^I$/$\epsilon_I$, transforming, in our conventions, in the ${\bf 8_s}$ of $SO(8)$. These can be decomposed under $SO(4)\times SO(4)\subset SO(8)$ to give parameters $\epsilon^a$/$\epsilon_a$ and $\epsilon^s$/$\epsilon_s$ transforming as  
${\bf (4,1)}$ and ${\bf (1,4)}$, respectively.

The non-trivially vanishing variations can then be written as
\begin{align}
\frac{1}{2}\delta\psi_\mu^a & = \,\left[ \nabla_\mu-\frac{1}{4}\left(\frac{\bar{z}\partial_\mu z - z \partial_\mu\bar{z}}{1-|z|^2}\right)\right]\epsilon^a+\frac{1}{2(1-|z|^2)^{1/2}}\Gamma_\mu \epsilon_a\,,\nn
\frac{1}{2}\delta\psi_\mu^s & =\, \left[ \nabla_\mu+\frac{1}{4}\left(\frac{\bar{z}\partial_\mu z - z \partial_\mu\bar{z}}{1-|z|^2}\right)\right]\epsilon^s+\frac{1}{2(1-|z|^2)^{1/2}}\Gamma_\mu \epsilon_s\,,
\end{align}
and
\begin{align}
\frac{1}{\sqrt{2}}\delta\chi^{abc} & = \,\frac{1}{1-|z|^2}\Big[\Gamma^\mu\partial_\mu\bar{z}\,\Omega^L\,_{abc}\,^d\epsilon_d-{\bar{z}}(1-|z|^2)^{1/2}\Omega^L\,_{abcd}\,\epsilon^d\Big]\,,\nn
\frac{1}{\sqrt{2}}\delta\chi^{stu} & = \,\frac{1}{1-|z|^2}\Big[\Gamma^\mu\partial_\mu z\,\Omega^R\,_{stu}\,^v\epsilon_v
-{z}(1-|z|^2)^{1/2}\Omega^R\,_{stuv}\,\epsilon^v\Big]\,,
\end{align}
together with their conjugate variations. Here we have used the $SO(4) \times SO(4)$ invariant tensors
\begin{equation}
\Omega^L = \frac{1}{4!}\Omega^L\,_{abcd}\,\dd x^a\wedge\dd x^b\wedge\dd x^c\wedge\dd x^d\,, \qquad \Omega^R = \frac{1}{4!}\Omega^R\,_{stuv}\, \dd x^s\wedge\dd x^t\wedge\dd x^u\wedge\dd x^v\,,
\end{equation}
where $a,b,c,d\in \{1,\dots,4\}$ and $s,t,u,v\in\{5,\dots, 8\}$ and $x^I$ are auxiliary coordinates on $\mathbb{R}^8$.

Noting the similarity with \eqref{Susyvars}, with $e^{\mathcal{-K}/2}=2^{3/2}(1-|z|^2)^{1/2}$ and $W=2^{5/2}$, we can utilise the results in the above subsection. 
In particular, the analysis for $\epsilon^s$ is the same, and we find on the corresponding {\it two component spinors}
$\sigma^{\hat{y}}\epsilon^s=-\epsilon^s$, $\epsilon_s=-e^{i\xi}\sigma^{\hat r}\epsilon^s$, 
$\epsilon^s=e^{A/2}e^{-i\xi/2}\eta^s_0$. For $\epsilon^a$ we find
$\sigma^{\hat{y}}\epsilon^a=\epsilon^a$, $\epsilon_a=-e^{-i\xi}\sigma^{\hat r}\epsilon^a$, 
$\epsilon^a=e^{A/2}e^{i\xi/2}\eta^a_0$. If we form the (four-component) 
Majorana spinors $\epsilon^a_{(M)}=\epsilon^a+\epsilon_a$ and $\epsilon^s_{(M)}=\epsilon^s+\epsilon_s$,
we conclude that the projections on the Killing spinors are given by
\begin{align}
\Gamma^{\hat{t}\hat{y}}\epsilon^a_{(M)}&= +\epsilon^a_{(M)}\,,\qquad\left[-\sin\xi\,\Gamma^{\hat{x}}+\cos \xi\,\Gamma^{\hat{r}}\right]\epsilon^a_{(M)} = -\epsilon^a_{(M)} \,,\nn
\Gamma^{\hat{t}\hat{y}}\epsilon^s_{(M)}&= -\epsilon^s_{(M)}\,,\qquad 
\left[-\sin\xi\,\Gamma^{\hat{x}}+\cos \xi\,\Gamma^{\hat{r}}\right]\epsilon^s_{(M)} = -\epsilon^s_{(M)}\,,
\end{align}
and hence solutions to the BPS equations preserve, generically, eight supersymmetries in
the $\mathcal{N}=8$ theory.

\section{Holographic one-point functions}
\label{sec:HolographicOnePointFunctions}
The holographic renormalization for the $SO(4)\times SO(4)$ invariant truncation of $\mathcal{N}=8$ gauged supergravity was discussed in
\cite{Gauntlett:2018vhk}, utilising the results of 
\cite{Cabo-Bizet:2017xdr}. Here we summarise and slightly extend that analysis. In this appendix we work
in Fefferman-Graham coordinates with
\begin{equation}
\dd s^2 = \gamma_{yy}\left(-\dd t^2 + \dd y^2 \right) + \gamma_{xx}\dd x^2 + \frac{\dd r^2}{r^2}\,, \qquad \mathrm{and}\qquad z = \mathcal{X} + i \mathcal{Y}\,.
\end{equation}
This is as in \eqref{Susyqansatz} with $N=1/r$ and we note that the coordinates $(r,x)$ used here are, in general, {\it not} the same
as in \eqref{Susyqansatz2} with $w=r+ix$.

Near the $AdS_4$ holographic boundary, at $r\to 0$, the equations of motion lead to the expansions
\begin{align}
-\gamma_{tt}=\gamma_{yy} & =  \Omega\, r^2 -\frac{1}{2}\Omega\left(X_1^2+Y_1^2 \right)+\frac{1}{8}\frac{\dot{\Omega}^2}{\Omega\Omega_x}+\frac{M}{r}+\ldots\,,\nonumber\\
\gamma_{xx} & = \Omega_x r^2 -\frac{1}{2}\Omega_x\left(X_1^2+Y_1^2 \right)-\frac{1}{4}\frac{\dot{\Omega}\dot{\Omega}_x}{\Omega\Omega_x}+\frac{1}{8\Omega^2}\left(4\Omega\ddot{\Omega}-3\dot{\Omega}^2\right)\label{eq:FGg}\nn
&\qquad -\frac{2}{3r}\frac{\Omega_x}{\Omega}\left(3M + 4\Omega X_1X_2 + 4\Omega Y_1 Y_2 \right)+\ldots\,,\nonumber\\
\mathcal{X} & = \frac{X_1}{r}+\frac{X_2}{r^2}+\ldots\,,\qquad
\mathcal{Y}  = \frac{Y_1}{r}+\frac{Y_2}{r^2}+\ldots\,,
\end{align}
where all the coefficients in the radial expansion are generically taken to be functions of $x$, and the ``dot" denotes a derivative with respect to $x$. Here we have allowed for a three-dimensional boundary
given by
\begin{equation}\label{eq:ds3}
\dd s_3^2 
= \Omega\left(-\dd t^2 +\dd y^2 \right) + \Omega_x \dd x^2\,,
\end{equation}
which is conformally flat.
Furthermore, there is a constraint between the coefficients in the above expansion which is given by
\begin{align}\label{eq:Mdot}
\dot{M} =-\frac{1}{6}\Big[ \frac{\dot{\Omega}}{\Omega}3 M+8\dot{\Omega}\left(X_1 X_2 + Y_1 Y_2 \right)+4\Omega \left( 2 X_2 \dot{X_1}+X_1\dot{X_2}+2Y_2 \dot{Y_1}+Y_1\dot{Y_2}\right)\Big]\,,
\end{align}
and is related to the Ward identity for the divergence of the stress tensor in the dual theory. 

For supersymmetric solutions satisfying the BPS equations \eqref{bpsderxi1} we have the additional constraints
 \begin{equation}
  M = -\frac{2}{3}\frac{\Omega}{\sqrt{\Omega_x}}\left(Y_1\dot{X}_1-X_1\dot{Y}_1 \right)\,,
  \end{equation}
  as well as
  \begin{equation}\label{eq:zSourcesVevsRelations}
  X_2 = -\frac{1}{2\Omega\sqrt{\Omega_x}}\left(Y_1\dot{\Omega}+2\Omega\dot{Y}_1 \right)\,,\qquad Y_2 = \frac{1}{2\Omega\sqrt{\Omega_x}}\left(X_1\dot{\Omega}+2\Omega\dot{X}_1 \right)\,,
  \end{equation}
 so that everything is fixed by $X_1$ and $Y_1$.
Furthermore, the function $\xi$ appearing in the Killing spinor has an expansion of the form 
  \begin{equation}
  \xi = -\frac{1}{r}\frac{\dot{\Omega}}{2\Omega\sqrt{\Omega_x}}+O(1/r^3)\,.
  \end{equation}

The boundary terms that we add to the bulk action \eqref{eq:Sgfirstone} are given by
\begin{equation}
S_{bdy} =\int  \dd^3 x\sqrt{-\gamma} \left[2K + 2\Big(\mathcal{W}-\frac{1}{2}R[\gamma] \Big)
+4\left(r\mathcal{X}\partial_r\mathcal{X}
+
{\mathcal{X}^2}\right)\right]\,.
\end{equation}
The first term is the standard Gibbons-Hawking term, the second is a counter term to remove divergences and the third term
ensures that the operators dual to $\mathcal{X}$, $\mathcal{Y}$ have conformal dimension $\Delta_{\mathcal{X}}=1$, $\Delta_{\mathcal{Y}}=2$, respectively, as required by supersymmetry\footnote{One can also consider finite 
counterterms of the form $\mathcal{X} R[\gamma]$ and $\mathcal{Y} R[\gamma]$ which would be associated with
different supersymmetric schemes \cite{Halmagyi:2017hmw,Cabo-Bizet:2017xdr}.}.
Calculating the stress tensor one finds 
\begin{align}\label{eq:vevF}
\langle \mathcal{T}^{tt} \rangle &= -\langle \mathcal{T}^{yy} \rangle  = -\frac{1}{\Omega^2}\left(3M+4\Omega Y_1Y_2\right) = -\frac{2}{\Omega^2\sqrt{\Omega_x}}\partial_x\left( \Omega X_1 Y_1\right)\,,\nn
\langle \mathcal{T}^{xx} \rangle & =-\frac{2}{\Omega\Omega_x}\left(3M +4 \Omega X_1 X_2 + 2\Omega Y_1Y_2\right) = \frac{2}{\Omega\Omega_x\sqrt{\Omega_x}}X_1\left(Y_1\dot{\Omega}+2\Omega\dot{Y}_1 \right)\,,
\end{align}
while the pseudo-scalar and scalar operator sources and VEVs are given by
\begin{align}
\mathcal{X}_s& = -4 X_2 = \frac{2}{\Omega\sqrt{\Omega_x}}\left(Y_1\dot{\Omega}+2\Omega\dot{Y}_1 \right)\,,\nn
\mathcal{Y}_s & = Y_1\,,\nn
\langle \mathcal{O}_{\mathcal{X}} \rangle & = X_1\,,\nn
\langle \mathcal{O}_{\mathcal{Y}} \rangle & = 4 Y_2  = \frac{2}{\Omega\sqrt{\Omega_x}}\left(X_1\dot{\Omega}+2\Omega\dot{X}_1 \right)\,.\label{eq:vevL}
\end{align}
In \eqref{eq:vevF} and \eqref{eq:vevL} the second equality impose the constraints arising from supersymmetry.
One can check that these  satisfy the Ward identities
\begin{align}\label{eq:Ward1text}
 \nabla_i\langle \mathcal{T}^i{}_{j}\rangle  = &\, \langle  \mathcal{O}_{\mathcal{X}}\rangle\partial_j\mathcal{X}_s+ \langle  \mathcal{O}_{\mathcal{Y}}\rangle\partial_j\mathcal{Y}_s\,,\nn
 \langle \mathcal{T}^i\,_i\rangle = & \, (3-\Delta_\mathcal{X})\, \langle \mathcal{O}_{\mathcal{X}}\rangle\mathcal{X}_s+(3-\Delta_\mathcal{Y})\, \langle  \mathcal{O}_{\mathcal{Y}}\rangle \mathcal{Y}_s\,.
\end{align}

Since the boundary theory has a time-like Killing vector, $\partial_t$, the total energy, $E$, is a conserved quantity. Using
$n^\mu=\Omega^{-1/2}(\partial_t)^\mu$, the unit normal to the surfaces of constant $t$ in the three-dimensional boundary metric, we have
\begin{equation}
E \equiv \int\dd x\,\dd y \sqrt{\Omega\Omega_x}n^\mu T_{\mu\nu}(\partial_t)^\nu = 
\int \dd x\, \dd y\, \Omega^2\sqrt{\Omega_x}\langle \mathcal{T}^{tt}\rangle \,.
\end{equation}
For the supersymmetric solutions this gives
\begin{equation}
E_{\mathrm{susy}} =-2\Delta y\int dx\partial_x\left(\Omega X_1 Y_1\right)\,,
\end{equation}
which demonstrates that the spatial average of the energy density vanishes for solutions periodic in $x$, or for 
those with sufficiently fast fall-off along the $x$ direction.

\section{Distributional sources}\label{sec:dist}
Equation \eqref{dwsz1} may be generalised to cases in which $z_1(\sigma)$ is a distribution. To that end, it is useful to consider the solution to the BPS equations
written in a gauge with an arbitrary $C(w)$ as in \eqref{Susyqansatz3}.
That is, we take the solutions with the metric transverse to the $t,y$ directions to be conformally flat, but we do not gauge fix as in \eqref{chalf}. Let $w_1, w_2$ be two points on the boundary $H=0$ curve and let $\gamma$ be an arbitrary path in the bulk region (with $H>0$) that connects these two points. Next consider the following integral along $\gamma$:
\begin{equation}
\int_\gamma 2 C(w) dw .
\end{equation}
This quantity has the following properties. First, it does not change as we deform the curve $\gamma$ into the bulk region since $C(w)$ is holomorphic (and assuming $C(w)$ is regular in the interior of the deformed region). Second, it is invariant under holomorphic coordinate transformations, $w\to w(\tilde w)$. Third, if the points $w_1, w_2$ are connected by a continuous segment of the boundary and $C(w)$ is regular on this segment then it is equal to $ i \int_{w_1}^{w_2} z_1(\sigma) d\sigma$. In the $C(w)=\frac{1}{2}$ gauge this is just the integrated version of \eqref{dwsz1}.

Now if $C(w)$ has a pole at some point on the boundary, then $z_1$ will be singular there. We may resolve this singularity by making a small deformation of the boundary around the pole, in which case the third
property above is satisfied for a choice of $w_1, w_2$ on the two sides of the pole. By shrinking this
deformation the third property should remain true even in the limiting case where the pole is on the boundary. In that case, however, computing this integral for two points $w_1, w_2$ that approach the pole may leave a finite contribution indicating the existence of a $\delta(\sigma)$ factor in $z_1(\sigma)$. Moving to the $C(w)=\frac{1}{2}$ gauge, such a pole would be mapped to infinity and the boundary will separate into two disconnected curves approaching infinity. Computing the above integral between end points on the two separate curves would then leave a finite contribution as the two end points approach infinity, corresponding to the $\delta(\sigma)$ contribution. An application of this to the Janus solution is discussed in appendix \ref{simpsol}.

\section{Simple explicit solutions}\label{simpsol}
In this section we present a number of simple analytic solutions to the BPS equations. The solutions all have the property
that $|z|<1$ in the relevant domain. 
As always, we
can multiply $z$ in the presented solutions by a constant phase, $e^{i\zeta}$, to get new solutions. 
In most cases we can obtain explicit expressions for $z_1(\s)$ and hence the one point functions
as given in section \ref{holoneptf}. We begin by discussing the Janus solution of \cite{DHoker:2009lky,Bobev:2013yra}.

\subsection{Janus revisited}
Recall the Janus solution that we discussed in section \ref{janusfirst}.
Along $r=0$ we have $\sigma=-\frac{c}{\sqrt{1+c^2}}e^{\sqrt{1+c^2}x}$, while along $r=\pi/c$ we can take
$\sigma=\frac{c}{\sqrt{1+c^2}}e^{\sqrt{1+c^2}x}$. The holographic boundary is then parametrised by $\sigma\in(-\infty,\infty)$, with
$\s=0$ associated with $x=-\infty$. For $\s\ne 0$ we can immediately determine the sources and VEVs in the dual theory 
with
\begin{align}
z_1=\frac{1}{\sqrt{1+c^2}}\frac{1}{\s}\,.
\end{align}
To determine what happens exactly at $\s=0$ requires some more work. 
We notice that we cross $\s=0$ as we move from $r=0$ to $r=\pi/c$ and hence $w=r+ix$ jumps by
$\pi/c$. From \eqref{dwsz1} this shows that we have a delta function in $z_1$ of the form $ -i\pi/c\delta(\s)$. Thus, 
we find\footnote{Note, in particular, we do not get 
$z_1=\frac{1}{\sqrt{1+c^2}}(\frac{1}{\s}-i\pi\delta(\s))$, as one may have expected from a naive
analytic continuation.}
\begin{align}\label{deetwo}
z_1=\frac{1}{\sqrt{1+c^2}}\frac{1}{\s}-i\pi\frac{1}{c}\delta(\s)\,.
\end{align}
We can make this argument more precise using the discussion in appendix
\ref{sec:dist}.
We first perform a coordinate transformation $ \tilde{w} = e^{-icw} $, so that $x \to -\infty$ is mapped to $\tilde{w}=0$ and the two boundaries are mapped to the positive and negative real axis of the $\tilde{w}$ plane. We then consider the integral $ \int 2 C(\tilde{w}) d\tilde{w}$ on a curve connecting the points $\tilde{w}=\pm e^{cx_0} $. From appendix
\ref{sec:dist} this integral is equal to $ i \int_{-\sigma_0}^{\sigma_0} z_1(\sigma) d\sigma $ where $\sigma_0=\frac{c}{\sqrt{1+c^2}}e^{\sqrt{1+c^2}x_0}$, but on the other hand it is also equal in the original $w$ coordinates to $ \Delta w = \frac{\pi}{c} $. Taking $x_0 \to -\infty $ we obtain the above $\delta(\sigma)$ factor. 

Using \eqref{deetwo}, we can then read off the sources and VEVs from section \ref{holoneptf}. In particular we notice that the sources are given by
$\mathcal{Y}_s=-\frac{\pi}{c}\delta(\s)$ and
$\mathcal{X}_s=4\frac{\pi}{c}\dot \delta(\s)$. If we consider the more general solutions with $z\to z e^{i\zeta}$
then we have, in particular, we then get $\mathcal{Y}_s=\frac{1}{\sqrt{1+c^2}}\frac{\sin\zeta}{\s}-\frac{\pi}{c}\cos\zeta \delta(\s)$ which 
is no longer purely a delta function.

\subsection{Half plane with zero at $x\to-\infty$}\label{UHP}

We take $0\le r< \infty$ and $-\infty\le x<\infty $ and consider
\begin{align}
H=\sinh(r\cos\alpha)e^{x\sin\alpha}\,,
\end{align}
where $\alpha$ is a constant.
This leads to the solution with
\begin{align}
ds^2&=\frac{1}{\sinh^2(r\cos\alpha)e^{2x\sin\alpha}}(-dt^2+dy^2)
+\frac{\cos^2\alpha}{\sinh^2(r\cos\alpha)}(dr^2+dx^2)\,,\nn
z&=-\frac{1}{\cos\alpha\coth(r\cos\alpha)-i\sin\alpha}\,.
\end{align}
We can now carry out the coordinate transformation as described in section \ref{identbdy}.
Explicitly, 
\begin{align}
\beta=H\,,\qquad \s=-\cot\alpha e^{x \sin\alpha}\,.
\end{align}
As we approach the $AdS$ boundary at $\beta=0$, we have $z\sim z_1\beta+\dots $ with
\begin{align}
z_1(\s)=\frac{\mathrm{cosec}\alpha}{\s}\,.
\end{align}
Note that the range of $\s \in (-\infty,0)$. It would be interesting to determine
whether there is a distributional source at $\s= 0$.

\subsection{Semi-infinite strip}
We take $0\le r\le \pi/c$, for some constant $c>0$
and $0\le x$ and consider
\begin{align}
H=\sin(c r)\sinh[\sqrt{1+c^2} x]\,.
\end{align}
This leads to the solution
\begin{align}
{}ds^2&=\frac{1}{\sin^2(c r)\sinh^2[\sqrt{1+c^2}x]}(-dt^2+dy^2)\nn
&\qquad+\left(c^2\cot^2(cr)+(1+c^2)\coth^2\left(\sqrt{1+c^2} x\right)-1\right)(dr^2+dx^2)\,,\nn
z&=-\frac{1}{c\cot(cr)-i(1+c^2)^{1/2}\coth\left(\sqrt{1+c^2} x\right)}\,.
\end{align}
The $AdS$ boundary has three components, region $I,II$ and $III$ with $r=0$, $x=0$ and $r=\pi/c$, respectively.
These are parametrised by a coordinate $\s$ in the range $\s\in(-\infty,-\frac{{\sqrt{1+c^2}}}{c}]$,
$\s\in[-\frac{{\sqrt{1+c^2}}}{c},\frac{{\sqrt{1+c^2}}}{c}]$ and
$\s\in[\frac{{\sqrt{1+c^2}}}{c},\infty)$, respectively, and the one point functions can be obtained from
\begin{align}
I:\qquad &z_1(\s)=-\frac{1}{\sqrt{1+c^2}}\frac{1}{\left[ \s +\frac{\sqrt{1+c^2}}{c} \right]^{1/2}}\frac{1}{\left[\s+\frac{1-c^2}{c\sqrt{1+c^2}}  \right]^{1/2}}
\,,\nn
II:\qquad &z_1(\s)=-i\frac{1}{c}\frac{1}{ \left(\frac{{1+c^2}}{c^2}-\s^2  \right)^{1/2} }
 \,,\nn
III:\qquad &z_1(\s)=
+\frac{1}{\sqrt{1+c^2}}\frac{1}{\left[ \s -\frac{\sqrt{1+c^2}}{c} \right]^{1/2}}\frac{1}{\left[\s-\frac{1-c^2}{c\sqrt{1+c^2}}  \right]^{1/2}}
\,.
\end{align}
In particular we observe the square root behaviour in $z_1$ at the two corners, as 
expected from the discussion in section \ref{cwcs}.

\subsection{Quadrant and wedge}
We take $0\le r< \infty$ and $0\le x<\infty $ and consider
\begin{align}
H=\sinh(r\cos\alpha)\sinh(x\sin\alpha)\,,
\end{align}
where $0< \alpha< \pi/2$ is a constant.
This leads to the solution with
\begin{align}
ds^2&=\frac{1}{\sinh^2(x\cos\alpha)\sinh^2(x\sin\alpha)}(-dt^2+dy^2)\nn
&\qquad
+\left[{\cos^2\alpha\coth^2(r\cos\alpha)+\sin^2\alpha\coth^2(x\sin\alpha)-1}\right](dr^2+dx^2)\,,\nn
z&=-\frac{1}{\cos\alpha\coth(r\cos\alpha)-i\sin\alpha\coth(x\sin\alpha)}\,.
\end{align}
The $AdS$ boundary has two components, region $I$ and $II$ with $r=0$ and $x=0$, respectively.
These are parametrised by a coordinate $\s$ in the range $\s\in(-\infty,0]$ and
$\s\in[0,\infty)$, respectively, and the one point functions can be obtained from
\begin{align}
I:\qquad &z_1=-\frac{1}{\sin\alpha}\frac{1}{\left[(\cot\alpha-\s)^2-\cot^2\alpha\right]^{1/2}}
\,,\nn
II:\qquad &z_1=-i\frac{1}{\cos\alpha}\frac{1}{\left[(\tan\alpha+\s)^2-\tan^2\alpha\right]^{1/2}}
 \,.
\end{align}

There is a generalisation of this solution to a wedge with opening angle $\pi/n$, for any integer $n$.
Using both polar coordinates $(\rho,\theta)$ and Cartesian coordinates $(r,x)$ 
we can take
\begin{equation}
H = \sum_{k=0}^{2n-1} (-1)^k e^{\rho \cos(\theta-\theta_k)} = \sum_{k=0}^{2n-1} (-1)^k e^{r \cos\theta_k + x\sin\theta_k},
\end{equation}
where $n$ is an integer, $\theta_k = \theta_0 + \frac{k \pi}{n}$ with $\theta_0$ an arbitrary angle.
It seems to satisfy $|z|<1$ everywhere in the bulk, but we have not proved this.

\subsection{Dipole}
The final solution we highlight lies in the upper half plane with $-\infty<r<\infty$, and $0\le x<\infty$. Unlike the previous example in section \ref{UHP}, here we place a delta function source with positive weight in the upper half plane and
a symmetrically placed one with negative weight in the lower half plane to ensure that the $x=0$ axis has $H=0$.
Explicitly we take
\begin{align}
H=K_0[\sqrt{r^2+(x-x_0)^2}]-K_0[\sqrt{r^2+(x-x_0)^2}]\,,
\end{align}
where $x_0>0$ is a constant. For this example, it is not straightforward to explicitly change coordinates from $(r,x)$ to
$(\beta,\s)$ to obtain the one point functions in closed form. A novel feature of this solution is that both $H\to 0$ and ${z}\to 1$
as one approaches infinity in the upper half plane, suggesting that there is a singularity on the holographic boundary, which could
be interesting to explore further.


\providecommand{\href}[2]{#2}\begingroup\raggedright\endgroup

\end{document}